\documentclass[]{article}
\usepackage{graphicx}
\makeindex
\begin{document}

\baselineskip 2.5ex
\parskip 3.5ex
\newcommand{\noi}{\noindent}
\renewcommand{\thesection}{\arabic{section}}
\renewcommand{\thesubsection}{\arabic{section}.\Alph{subsection}}
\renewcommand{\theequation}{\arabic{section}.\arabic{equation}}
\renewcommand{\a}{\alpha}
\renewcommand{\b}{\beta}
\newcommand{\e}{\epsilon}
\newcommand{\g}{\gamma}
\newcommand{\G}{\Gamma}
\renewcommand{\d}{\delta}
\newcommand{\D}{\Delta}
\renewcommand{\vec}{\bm}
\newcommand{\svec}[1]{\mbox{{\footnotesize $\bm{#1}$}}}
\newcommand{\bm}[1]{\mbox{\boldmath $#1$}}
\newcommand{\sumc}[3]{\sum_{#1=#2}^{#3}}
\newcommand{\pder}[2]{\frac{\partial {#1}}{\partial {#2}}}
\newcommand{\pdert}[2]{\frac{\partial^2 {#1}}{\partial {#2}^2}}
\newcommand{\fder}[2]{\frac{\delta {#1}}{\delta {#2}}}
\newcommand{\PDD}[3]{\left.\frac{\partial^{2}{#1}}{\partial{#2}^{2}}\right|_
{#3}}
\newcommand{\PD}[3]{\left.\frac{\partial{#1}}{\partial{#2}}\right|_{#3}}
\newcommand{\der}[2]{\frac{d {#1}}{d {#2}}}
\newcommand{\diff}[1]{{\rm d}{#1}}
\newcommand{\fif}{\mbox{$\Longleftrightarrow$}}
\newcommand{\hsk}{\hspace{2em}}
\newcommand{\beq}{\begin{equation}}
\newcommand{\bequo}{\begin{quotation}}
\newcommand{\beqa}{\begin{eqnarray}}
\newcommand{\eeq}{\end{equation}}
\newcommand{\equo}{\end{quotation}}
\newcommand{\eeqa}{\end{eqnarray}}
\newcommand{\la}{\langle}
\newcommand{\ra}{\rangle}
\newcommand{\hsp}{\hspace*{8em}}
\newcommand{\rate}[2]{I_{\tilde{#1}}^{(#2)}}
\newcommand{\lima}[1]{\lim_{#1 \rightarrow \infty} \frac{1}{#1} \ln}
\newcommand{\meas}[2]{{#1}(\mbox{d} {#2})}
\newcommand{\limin}[1]{\lim_{#1 \rightarrow \infty}}
\newcommand{\un}{\underline}
\newcommand{\dd}{\mbox{d}}
\newcommand{\lb}{\label}
\newcommand{\fr}[1]{(\ref{#1})}
\newcommand{\non}{\nonumber}
\newcommand{\bmc}[1]{\mbox{\boldmath $#1$}^{C}}
\newcommand{\bml}[1]{\mbox{\boldmath $#1$}^{L}}
\newcommand{\sbm}[1]{\small{\mbox{\boldmath $#1$}}}
\newcommand{\fns}{\footnotesize}
\newcommand{\cvec}[3]{\left( \begin{array}{c} #1\\#2\\#3 \end{array} \right)}
\newcommand{\matr}[9]{\left(\begin{array}{ccc} #1&#2&#3\\#4&#5&#6\\#7&#8&#9
\end{array} \right)}
\newcommand{\see}{$\rightarrow$}
\newcommand{\map}{\rightarrow}
\newcommand{\maps}{\rightarrow}
\newcommand{\imply}{\Rightarrow}
\newcommand{\implies}{\Rightarrow}
\newcommand{\define}{\,\raisebox{-.4ex}{$\leftharpoondown$}{\hspace{-1.0em}\
raisebox{.45ex}{=\hspace{-.6em}=}\,}}
\newcommand{\defines}{\,\raisebox{-.4ex}{$\leftharpoondown$}{\hspace{-1.0em}
\raisebox{.45ex}{=\hspace{-.6em}=}\,}}
\newcommand{\bra}[1]{\langle{#1}|}
\newcommand{\ket}[1]{|{#1}\rangle}
\renewcommand{\i}[1]{\index{#1}}

\newcommand{\be}{\begin{equation}}
\newcommand{\benn}{\begin{eqnarray*}}
\newcommand{\eenn}{\end{eqnarray*}}
\newcommand{\ba}{\begin{array}}
\newcommand{\bea}{\begin{eqnarray}}
\newcommand{\ee}{\end{equation}}
\newcommand{\ea}{\end{array}}
\newcommand{\eea}{\end{eqnarray}}
\def\real{\hbox{\rm\setbox1=\hbox{I}\copy1\kern-.45\wd1 R}}
\def\nonn{\hbox{\rm\setbox1=\hbox{I}\copy1\kern-.45\wd1 N}}
\def\forc{\hbox{\rm\setbox1=\hbox{I}\copy1\kern-.45\wd1 F}}
\def\flux{\hbox{\setbox1=\hbox{f}\copy1\kern-.45\wd1 f}}
\newcommand{\al}{\alpha}
\newcommand{\bet}{\beta}
\newcommand{\gam}{\gamma}
\newcommand{\lam}{\lambda}
\newcommand{\eps}{\epsilon}
\newcommand{\ov}{\overline}
\newcommand{\Lrar}{\Leftrightarrow}
\newcommand{\rar}{\rightarrow}

\newtheorem{entry}{}[section]
\newcommand{\bent}[1]{\vspace*{-2cm}\hspace*{-1cm}\begin{entry}\lb{e{#1}}\rm}
\newcommand{\eent}{\end{entry}}
\newcommand{\fre}[1]{{\bf\ref{e{#1}}}}

\begin{center}
{\Large
Relational patterns of gene expression via nonmetric multidimensional
scaling analysis}
\end{center}

running head:
Relational patterns via nonmetric MDS

Y.-h. Taguchi$^{1,2,*}$ and  Y. Oono$^{3,4}$,

\noi
$^1$Department of Physics, Faculty of Science and Technology,
Chuo University, 1-13-27 Kasuga, Bunkyo-ku, Tokyo 112-8551, Japan,\\
$^2$Institute for Science and Technology,
Chuo University, 1-13-27 Kasuga, Bunkyo-ku, Tokyo 112-8551, Japan,\\
and\\
$^3$Department of Physics,  1110 W. Green Street,
              University of Illinois at Urbana-Champaign,
              Urbana, IL 61801-3080, USA.\\
$^4$Institute for Genomic Biology,
              University of Illinois at Urbana-Champaign,
              Urbana, IL 61801, USA.\\

\medskip

\hrule
\noi
* To whom correspondence should be addressed

\pagebreak

\twocolumn
\noi
{\bf ABSTRACT}\\
{\bf Motivation}: Microarray experiments result in large scale
data sets that require extensive mining and refining to extract
useful information. We demonstrate the usefulness of (nonmetric)
multidimensional scaling (MDS) method in analyzing a
large number
of genes. Applying MDS to the microarray data is certainly not
new, but the existing works are all on small numbers
  ($<$ 100) of
points to be analyzed.  We have been developing an efficient novel
algorithm for nonmetric multidimensional scaling (nMDS) analysis
for very large data sets as a maximally unsupervised data mining
device. We wish to demonstrate its usefulness in the context of
bioinformatics (unraveling relational patterns among genes from
time series data in this paper).
\\
{\bf Results}:
The Pearson correlation coefficient with its sign flipped is used  to
measure the dissimilarity of the gene activities in transcriptional
response of cell-cycle-synchronized human fibroblasts to serum
[Iyer {\it et al}., Science {\bf 283}, 83 (1999)]. These
dissimilarity data have been analyzed with our nMDS algorithm to
produce an almost circular relational pattern of the genes.  The
obtained pattern expresses a temporal order in the data in this
example; the temporal expression pattern of the genes rotates
along this circular arrangement and
is related to the cell cycle.
For the data we analyze in this paper we observe the following.
If an appropriate preparation procedure
is applied to the
original data set, linear methods such as the principal component
analysis (PCA) could achieve reasonable results, but without data
preprocessing linear methods such as PCA cannot achieve a useful
picture.  Furthermore, even with an appropriate data
preprocessing, the outcomes of linear procedures are not as
clearcut as those by nMDS without preprocessing.
\\
\noi
{\bf Availability:}
The fortran source code of the method used in this analysis  (`pure
nMDS')  is available at \\  http://www.granular.com/MDS/\\
\noi
{\bf Contact:}
tag@granular.com, yoono@uiuc.edu\\

\noi
{\bf INTRODUCTION}\\
Each DNA microarray experiment can give us information about
the relative populations of mRNAs for thousands of genes.   This
implies  that without extensive data mining it is often hard to
recognize any useful information  from the experimental results.
In this paper we demonstrate that a nonmetric multidimensional
scaling (nMDS) method can be a powerful unsupervised means to
extract relational patterns in gene expression.  A data mining
procedure may be useful, if it is flexible enough to incorporate
any level of supervision, but we believe that the most basic
feature required for any good data mining method is to be able to
extract recognizable significant patterns reproducibly without
supervision. In this sense our nMDS method is clearly
demonstrated to be a useful means of data mining.

Since MDS is a very natural tool to visualize the salient features
in the data in general, initially we expected many existing
published works but to our knowledge actually there are not so
many.  Besides, these existing examples (only a few published
papers in the bioinformatics area, {\em e.g.}, Dyrskjot
{\it et al.}\
(2002)),
are mostly metric MDS examples and the number of points to be
imbedded is small (about 30).  Shmulevich and Zhang (2002)
applies a nonmetric MDS method,
but again they classify tumor types.  That is, they study
less than 100 points to embed (visualize).

We have been developing an efficient nMDS technique for large
data sets (Taguchi and Oono, 1999, Taguchi {\it et al}., 2001,
Taguchi and Oono, 2004).  The
input is the rank order of (dis)similarities among the objects (in
the present case, genes). Our algorithm is  maximally nonmetric
in the sense that any introduction of intermediate metric
coordinates obtained by monotone regression common to the
conventional nMDS methods is avoided.

Data compression is essentially a problem of linear functional
analysis as Donoho {\it et al}.\ (1998) stresses. In contrast, we
believe that nonlinear methods should be important to
data-mining.  There are linear algebraic methods such as the
principal component analysis (PCA) for data mining, but it is
expected that nonlinear methods are, in principle, more powerful
than linear methods. The present paper illustrates this point.
Indeed, in our case PCA cannot find any comprehensible
pattern in low dimensional spaces without an appropriate data polishing.

\noi
{\bf SYSTEMS AND METHODS}\\
\noi
{\bf Systems to Analyze}\\
The gene activities in transcriptional response of cell
cycle-synchronized human fibroblasts to serum reported by Iyer
{\it et al}.\ (1999) are analyzed.  The microarray data used in this
analysis is available at
http://genome-www.stanford.edu/serum/fig2data.txt.\\

\noi
{\bf Other Possible Analysis Methods}\\
To extract interpretable patterns from microarray  data, cluster
and linear multivariate analyses seem to be the two major
strategies. However, these methods may not always work as
shown here in the example of the analysis of time series data.

The cluster analysis seems to be the most common
approach  (Slonim, 2002). For example, the hierarchical clustering
method seems to be popular (Eisen {\it et al}., 1998). Classifying
the expression patterns as functions of time is often attempted
by clustering methods (Spellman {\it et al}., 1998).
However, it is
difficult to see  continuous relations among genes by cluster
analysis, because it must classify genes into a limited number of
categories.  For example, it is difficult to see how genes are
expressed temporally. In contrast to clustering analyses, nMDS
can place genes in a continuous space, so we can easily recognize
the relationship among genes that change in a continuous manner.
Thus, the nMDS method can capture features which cluster
analysis is hard to find.

The principal component analysis (PCA) is a method that
visualizes the relationship among genes in a continuous vector
space. The main idea is to choose a data-adapted basis set, and to
make a subspace that can capture salient features of the original
data set.  In principle, the method could capture  the
continuous relations in the gene expression pattern, but the dimension of the
subspace may not be low even if the data are on a very low
dimensional manifold.  In short, the information compression
capability of linear methods is generally weak.  This can be
illustrated well by the data we wish to analyze in this paper: PCA
does not capture the temporal order as clearly as nMDS. Figure 1
exhibits the two dimensional space spanned by the first two
principal components. The temporal expression pattern is hardly
recognized in the result. This should be compared with the nMDS
results in
Fig. 4.

\marginparsep -9mm
\marginpar{\hspace{10cm}\fbox{Fig. 1}}
\marginparsep 4mm

Apparently, however,
some linear analysis methods work well.  Holter {\it et al}.\ (2000)
demonstrated that
the singular value
decomposition (SVD; a linear method) is remarkably successful in
extracting the characteristic modes.   The reader must wonder
why there is a difference between this result and the one due to
PCA that is not successful.  The secret is in the highly nonlinear
`polishing' of the original data (proposed by Eisen {\it et al}. (1998)).
However, the role of this  nonlinear polishing must be considered
carefully, because it can generate a spurious temporal
behavior.
Therefore, we relegate the comparison of these linear methods
with data preprocessing and our nMDS to Appendix II.  The salient
conclusions are:\\
(1) Linear methods such as
PCA and SVD could perhaps achieve reasonable results,
if a data preparation scheme is chosen appropriately.
However, even the best linear results are generally fairly inferior
to nonlinear
results.\\
(2) The data preparation such as the `polishing' used by Holter {\it et
al.}\  (2000) could actually corrupt the original data (as illustrated in
Appendix II), and should be avoided.

An interesting proposal to extract a temporal order is to use the
partial least squares (PLS)
regression (Johansson {\it et al}., 2003). In this case one may
assume a temporal order one wishes to extract (say, a sinusoidal
change in time), and the original data are organized around the
expected pattern.  This is, so to speak, to analyze
a set of data according to a certain prejudice.  Although during the process of
organization no supervision is needed, the pattern to be extracted
(that is, the `prejudice') must be presupposed.
Furthermore, if a clear objective pattern could be extractable by
this method, certainly nMDS can achieve the same goal without
any presupposed pattern required by PLS.

Metric multidimensional scaling methods (MDS) may also be used, but
they depend on the definition of the dissimilarity. Therefore, unless the
measure of the
dissimilarity is (almost) dictated by the data or by the
context of the data analysis, arbitrary elements are introduced.
Even if a dissimilarity measure is already given, if the measure
is not positive definite (as in the case of the microarray data), the
metric MDS requires its conversion to a positive definite metric.
Thus, an arbitrary factor
may further be introduced, and the metric supposedly
capturing detailed information could carry spurious information
(disinformation, so to speak) as well. Indeed, this extra factor can
completely change the possible geometrical features in the data
as can be seen from the following simple example.  Take three
samples A, B and  C with the dissimilarities given as AB $=-1,$
BC $=0$, AC $= 1$.  To convert them into positive numbers we
shift the origin.  If we add 2, A, B, C can be arranged consistently
in 1D.  If we add 1.5, the triangle inequality is violated.  If we add
3, it can be geometrically realized but not in 1D.
nMDS is completely free from this problem, because adding a common
number cannot change the ordering of the dissimilarities. Another
disadvantage of metric MDS (and of linear multivariate methods) is
that these methods are vulnerable to missing or grossly inaccurate data.

The application of nonmetric MDS is a very natural idea, but there are
only a small number of published examples. Furthermore, to our
knowledge there is no paper analyzing large number of objects by
nMDS. Dyrskjot {\it et al.}\ (2002) is a rare example of applying nMDS,
but, in contrast to ours, they never used MDS to analyze genes; they
analyzed 40 tumors with a package in SPSS.  Our main purpose is to
extract information on genes themselves.

The cluster analysis with the aid of self-organizing maps (SOM)
is definitely a nonlinear data analysis method, but as we have
seen in Kasturi {\it et al}.\ (2003) it is not generally suitable for
extracting temporal order.  Kasturi {\it et al}.\ comment that SOM
is not particularly better than the ordinary cluster analysis.
Besides, as can be seen from the fact that the use of a particular
initial condition can be a methodological paper (Kanaya {\em et
al}., 2001), we must worry about the {\em ad hoc}
initial-condition dependence of the results. \\

\noi
{\bf ALGORITHM}\\
\noi
{\bf Basic idea of algorithm for nMDS}\\
The philosophy of nMDS  (Shepard 1962a, b,
Kruskal 1964a,b) is to find a constellation in a certain metric space ${\cal
R}$ of points representing the objects under study (genes in the
present case) such that the pairwise distances $d$ of  the points
in ${\cal R}$ have the rank order in closest agreement with the
rank order of the pairwise dissimilarities $\d$  of the
corresponding objects that are given as the raw (or the original)
input data.

The conventional nMDS methods assume a certain intermediate
pair distance  $\hat{d}$ that is chosen  as close as possible to $d$
for a given object pair under the condition that it is monotone
with respect to the actually given ordering of the dissimilarities
$\d$.  The choice of $\hat{d}$ is not unique.  The discrepancy
between $d$ and $\hat{d}$ is called the stress, and all the
algorithms attempt to minimize it. Depending on the choice of
$\hat{d}$ and on the interpretation of ``as close as,'' different
methods have been proposed (see, for example, Green {\it et al}.
(1970), Cox and Cox (1994), and Borg and Groenen (1997) ). The
choice of $\hat{d}$ affects the outcome.  $\hat{d}$ is required
only by technical reasons for implementation of the basic
nonmetric idea, so to be faithful to the original idea due to
Shepard (1962a, b) we must compare $\d$ with $d$ directly.
Our motivation is to make an algorithm that is maximally
nonmetric in the sense that we get rid of $\hat{d}$.

The basic idea of this `purely nonmetric' algorithm is as follows
(Taguchi and Oono, 1999, Taguchi {\it et al}., 2001, Taguchi and
Oono, 2004): in a metric space ${\cal R}$ (in this paper,
$D$-dimensional Euclidean space ${\bm R}^D$ is used) $N$ points
representing the $N$ objects are placed as an initial
configuration. For this initial trial configuration we compute the
pair distances $d(i,j)$, and then rank them according to their
magnitudes. Comparing this ranking and that according to the
dissimilarity data $\delta(i,j)$, we compute an appropriate `force'
that moves the points in ${\cal R}$ to reduce the discrepancy
between these two rankings.  After moving the points according
to the `forces', the new `forces' are computed again, and the whole
adjusting process of the object positions in ${\cal R}$ is iterated
until they converge sufficiently.  The details are in Appendix I.

nMDS can usually recover geometrical objects correctly (up to
scaling, orientation, and direction) when there are sufficiently
many (say, $\ge 30$) objects.
Furthermore, even with metric data converting them into the
corresponding rank order data may be a general strategy to extract
robust features in the metric data.
Therefore, nMDS is a versatile
multivariate analysis method.

It is desirable to have a criterion for convergence (analogous to
the level of the stress in the conventional nMDS), or a measure of
goodness of embedding.  To this end let us recall the Kendall
statistics $K$ (p364, Hollander and Wolfe (1999)),
\[
K = \sum_{\la\a,\b\ra} \mbox{sign}[(d_\a-d_\b)(\d_\a-\d_\b)],
\]
where the summation is over all the pairs of dissimilarities
(distances between objects) $\la \a, \b\ra$ (i.e., $\a$ (also $\b$)
denotes  a pair of objects). Usually, this is used for a statistical
test to reject the null hypothesis that $\{d(i,j)\}$ does not
correlate with $\{\d(i,j)\}$ (The contribution of ties is usually negligible
for large data set, so we do not pay any particular
attention to tie data).

Here, we use this value to estimate the number of the objects
embedded correctly.
If $N'$ objects are correctly embedded,
and if we may assume that the rest are uncorrelated, then $K$ is
expected to be
\[
n'(n'-1)/2 + O[N^2]> K > n'(n'-1)/2 - O[N^2],
\]
where $n' = N'(N'-1)/2$. This can be shown as follows.  Since $N'$
is correctly embedded, $n'$ pairs must be correctly ordered, so
the contribution from these correct points to $K$ is $n'$.  We
assume the rest is random.  Their contribution is the random sum
bounded by $\sum_{\la \a,\b\ra} \pm 1$ that is of order
$\sqrt{_{_{N}C_{2}}C_{2}}\simeq O[N^2]$, if we assume that the
central limit theorem applies. This gives the latitude in the above
inequalities. If the embedding is successful for the majority of
the objects, then $n' = O[N^2]$, so we may ignore the contribution
of the bad points. Thus, we may estimate
\[
N' \simeq \sqrt{2\sqrt{2K}}.
\]
Therefore, we adopt $100\sqrt{2\sqrt{2K}}/N$\%  as an indicator
of the goodness of embedding.   Although one might criticize that
this is a crude measure of goodness, notice that MDS does not
have any clear criterion of the goodness of embedding except for
the so-called stress.
In contrast to the stress, $N'$  gives an effective number
of
correctly embedded points.

The plausibility of the above estimate of the number of correctly
embedded points may be
illustrated as follows: We prepare a data set of 200 objects whose
subset $A$ consisting of $M$ objects is embeddable in a 3D
space, but whose remaining $200-M$ objects are not.  The subset $A$
consists of $M$ randomly positioned points in 3D; their $x,y,z$
coordinates are uniform random numbers
in $[-1,1]$.
The dissimilarity  between
objects $i,j \in A$ is defined by
$$
\delta_{ij} \equiv \sqrt{\frac{2}{3}[(x_i-x_j)^2+(y_i-y_j)^2+(z_i-z_j)^2]}.
$$
The dissimilarities between the points both of which are not in
$A$ are chosen to be the square root of random numbers uniformly
distributed in $[0,2]$.
With these choices, the variance of the dissimilarities are unity
for both cases.
We
embedded these 200 objects into a 3D space using these
dissimilarities for several values of $M$.

\marginparsep -9mm
\marginpar{\hspace{10cm}\fbox{Table. 1}}
\marginparsep 4mm

As can be seen from the results in Table 1, $N'$ monotonically
increases with $M$.
The difference between them decreases as $N-M$ decreases.
Actually, $N'-M \propto \sqrt{N-M}$.  Thus, we can expect
that $N'/N$ is asymptotically an effective fraction of correctly
embedded points
in the $N \rightarrow \infty$ limit with $(N-M)/N $ kept constant.

We must also discuss the initial configuration dependence of the
result. Our algorithm is not free from the problem of local minima
as all of the previously proposed algorithms for nMDS and as high
dimensional nonlinear optimization problems in general. However,
generally speaking, this dependence has only a very minor effect.
This has been checked for the
fibroblast data (see below).

\noi
{\bf RESULTS}\\
We have found that the fibroblast data may be embedded in a two
dimensional space approximately as a ring (Fig.\ 2). The estimated
number of correctly embedded genes is about 480 among all the
517 genes (i.e., the goodness of embedding is more than 90\%).
Also 516 out of 517 genes have $P < 0.005$ confidence level (see
Appendix I). One might wonder that this is too lax a criterion,
because we should apply a multiple comparison criterion when we
have so many  a
number of genes. However,  the $P$-value is still
very small: we
still have 515 genes with the $P<0.005$ confidence level of correct
embedding under multiple comparison criterion.  That is, we need not
change the gross upper bound estimate of $P$.

\marginparsep -9mm
\marginpar{\hspace{10cm}\fbox{Fig. 2}}
\marginparsep 4mm

The obtained configuration is insensitive to the initial
configurations. To see this we constructed two 2D embedding
results starting from two different random initial configurations.
With the aid of the Procrustean similarity transformation (Borg
and Groenen, 1997) one result is fit to the other (notice that our
procedure is nonmetric, so to compare two independent results,
appropriate scales, orientations, etc., must be optimally chosen).
Fig.\ 3 demonstrates the close agreements of $x$- and
$y$-coordinates of the two results.  As illustrated, the
dependence on the initial conditions is very weak, and we may
regard the embedded structure as a faithful representation of the
information in the original data. Thus, we  conclude that the
obtained configuration is sufficiently reliable.

\marginpar{\hspace{10cm}\fbox{Fig. 3}}

\marginpar{\fbox{Fig. 4}}

This ring-like arrangement of the genes faithfully represents the
temporal expression patterns of the genes as can be seen clearly
from the rotation of the expression peaks around the ring (Fig.\
4a).  It is noteworthy that the angle coordinate assigned to the
genes according to the result shown in Fig.\ 2 automatically gives
the figure usually obtained only through detailed Fourier analysis
(Fig.\ 4b). These figures attest to the
usefulness of nMDS, a nonlinear data mining method, for
extracting nontrivial patterns in large scale  data sets.

One might criticize that the temporal pattern just mentioned is
rather trivial and can be expected intuitively, because the temporal
pattern may be recognized, if we pay attention to the peaks only.
However, the obtained ordering of the genes along the ring (i.e.,
the angle coordinate) cannot simply be obtained by ordering genes
according to  their peak times alone.  There are genes that exhibit
peaks only at one time, so there is no way to order these genes
sharing the peak time (just as there is no way to order genes in
one cluster in cluster analyses without inspection).  The detailed
ordering obtained by nMDS reflects biologically meaningful events
as discussed below.

Most genes directly related to cell cycle have peaks at 24 hr.  All
of them are tabulated in Table 2; the information of the genes
were obtained with the aid of AceView
(http://www.ncbi.nlm.nih.gov/IEB/Research/Acembly/).
Thus, we seem to observe the transition from G2 to G1 through M
phase along the ring between angle variables 1.8 and 3 (needless
to say,
the angle coordinate difference of order  0.1 is not
significant). That is, the ordering found by nMDS is consistent
with
some known functions of genes. Since the starved fibroblasts
are supposedly in G1, perhaps one full cell cycle may have been
captured in the original data. This may explain why the peak
positions make one complete rotation. One might question that
this full rotation is simply due to the use of all the data.  Even if
we use the data only up to 16hr from the start, we can actually
almost reproduce the gene configuration exhibited in Fig.\ 2.
Obviously, the peaks up to time 16 hr cannot complete any full
circle around it.  Thus, the observed full rotation in Fig 4a
is not an
artifact of our data analysis.
Thus, we may conclude that nMDS captures, as cluster analyses do,
major expression patterns of the genes but with a more natural time
ordering in the current example.

\marginparsep -9mm
\marginpar{\hspace{10cm}\fbox{Table. 2}}
\marginparsep 4mm

Although there is a recent paper claiming that the data we are
analyzing can provide information about the gene ontology (Laegreid
{\it et al}. (2003)), we do not think the data can be used for this
purpose  (Appendix
III).

As has been clearly demonstrated,  the 2D embedding is
statistically natural and biologically informative.  Still the 2D
embedding is not perfect, so it is interesting to see what we
might obtain by `unfolding' the 2D data, adding one more axis.   The
unfolded result is shown in Fig.\ 5. Here, the angular coordinates
$\phi$ and $\theta$ of the spherical coordinate system is
determined by the $xy$-plane whose $x$-(resp., $y$-)axis is the
first (resp., the second) principal component of the 3D embedded
result.  The total contribution of these two components is 86\%.
We do not recognize any clear pattern other than that captured in
the 2D space.  Therefore, we may conclude that the 2D embedding
result is sufficiently reliable and informative.

\marginparsep -9mm
\marginpar{\hspace{10cm}\fbox{Fig. 5}}
\marginparsep 4mm

However, one might wish to be more quantitative.  Traditionally
within the multidimensional scaling methods, there is no very
clear way to determine the dimensionality of the embedding
space.  Here, we propose a method utilizing the correlation among
principal coordinates.  Basically, the strategy is as follows.
Suppose we embed identical data into $n$-dimensional and
$(n+1)$-dimensional spaces.  Using the embedded results, we can
construct principal axes with the aid of PCA.  Then, we study the
correlation coefficients of the coordinates of the points.  It is
usually the case that the first $n$ principal axes of the
$(n+1)$-dimensional embedding result agree well (have high
correlation coefficients) with the $n$ principal axes of the
$n$-dimensional embedding result.  Now, we see the correlation
for the $(n+1)$th axis of the $(n+1)$-dimensional embedding result
and $n$ principal axes obtained from the $n$-dimensional
embedding.  If the correlations are low enough, we may say
$n$-dimensional space captures the main features in the data.  We
apply this to $n=2$. In the following table nDX denotes the Xth
principal axis for the $n$-dimensional embedding (Table 3).
As can be seen from the table, we may conclude that 2D is  enough
to capture the main features in the data.  We could even devise a
statistical test based  on the correlation coefficients, but we do
not go into this further in this paper.

\marginpar{\fbox{Table 3}}

In this paper we have demonstrated that nMDS has a capability
of extracting a significant pattern without any supervision or
preprocessing.   However, the data we have analyzed is a very high
quality well structured data.  A natural question is what nMDS can
offer for less well structured data.

As an example, we here show the nMDS analysis result of
Cho {\it et al}.\
(2001) on the cell cycle of human foreskin cells.
Their data were analyzed in detail by Shedden and
Cooper (2002).  According to
their analysis periodicity of the genes in Cho
{\it et al.}\
is statistically insignificant; the data randomly permuted along
the time axis
exhibit periodic patterns of similar or greater strength than the
original experimental data. As was done by Shedden and Cooper, we
have selected 300 periodic genes from both original data and
randomly time-reordered data.
One should keep in mind that it was impossible to distinguish these
two data by inspection.
The two upper figures in Fig.\ 6 is
the results when we applied our nMDS to these two data sets.

\marginparsep -9mm
\marginpar{\fbox{Fig. 6}}
\marginparsep 4mm

First, one can notice that both of them
exhibit ring-like structures. It is natural because periodic genes 
are selected.
However, in contrast to the analysis by Shedden and Cooper, we
can detect some differences between these two figures. Along the
ring-like structure in the original data, the distribution of the
genes is not uniform and has two dense regions placed
diametrically opposite to each other. On the other hand, such a
structure cannot be seen in the randomized data.
Needless to say, the forced periodicity by an artificial selection
of the data is detected in both cases as a ring structure, so one
might conclude, as Shedden and Cooper did, the observed
periodicity  is not statistically significant.  However, the
difference between the original data and the randomized data
clearly tells us the conclusion is premature.

To highlight the capability of nMDS, the same selected data were
analyzed by PCA with the preprocessing discussed in Appendix II
(the data standardization by Method II) (the two
middle figures in
Fig.\ 6).  We see much more diffuse rings than obtained by nMDS.
It is however interesting to note that PCA captures the
nonuniformity of the distribution of genes noted in the above.

Thus, we may conclude that nMDS has a capability of detecting subtle
differences.  In the example here, this could be detected by PCA as
well (with normalization), but it is clear that nMDS is much more
crisp than PCA. However, the data analyzed here are not really the
original data, but a subset of the original data selected by the
good correlation to sinusoidal time dependence.
Therefore, as duly criticized by Shedden and Cooper, the periodicity
in the analyzed data can largely be an artifact.

Therefore, to avoid this possible artifact we made a 300 gene
subset consisting of genes with largest expression level
variations to avoid possible noise effects. The 2D embedding
result of these genes with the aid of nMDS is shown in the bottom
left of Fig.\ 6.
Although diffuse, a ring-like structure is still visible.  More
interestingly, two distribution peaks diametrically placed on the
ring are clearly visible.  If we randomly permuted the same data
along the time axis, the embedded result (bottom right) loses both
the ring-like structure and the peaks.  Therefore, we may conclude that
the original data contains some information about the cell-cycle related
gene expression contrary to the conservative conclusion by Shedden and
Cooper.  We have no intention to criticize their study that was admirably
critical and careful; we simply wish to  demonstrate
that nMDS can detect subtle structures.

Iyer {\it et al}.\
selected the genes to analyze according to the significance
of the changes in their expression levels, so the temporal
structure detected by our nMDS analysis is not an artifact of the
methodology in contrast to that exhibited in Cho {\it et al}.\
(2001).

\noi{\bf CONCLUSION}\\
We have demonstrated that the nMDS can be a useful tool for data
mining.  It is unsupervised, and perhaps maximally nonlinear.  Our
algorithm is probably the simplest among the nonmetric MDS
algorithms and is efficient enough to enable the analysis of a few
thousand objects with a small laptop machine.

The nMDS algorithm works on the binary relations among the
objects, so if there are $N$ objects, computational complexity is
of order $N^2$ at least.  Therefore, it is far slower than linear
methods such as PCA, although our nonlinear algorithm is
practically fast enough; we have used for this work a
small laptop machine.
(Mobile Celeron 650MHz cpu with 256MB RAM).
Typically, the necessary number of iteration is about a hundred,
and it takes only a few minutes for a few hundreds of genes. As
has been pointed out and illustrated, with an appropriate data
preprocessing a certain linear method could give us a reasonable
result with less computational efforts.  Although in this paper we
have not made any particular effort to reduce computational
requirements, a practical way to use nMDS might be to prepare an
initial configuration by a linear method with an appropriate data
preprocessing method that is verified to be consistent with the
full nMDS results.

\noi
{\bf APPENDIX I}\\
\noi
{\bf `Purely' non-metric MDS algorithm}\\
Suppose $d(i,j)$ is the distance between objects $i$ and $j$ in
${\cal R}$.  Let the ranking of $\d(i,j)$ among all the input
dissimilarity data be $n$ and that of $d(i,j)$ among all the
distances between embedded pairs be $T_n$. If $n>T_n$ (resp., $n
< T_n$), we wish to `push' the pair $i$ and $j$ farther apart (resp.,
closer) in ${\cal R}$. Intuitively speaking, to this end we
introduce an `overdamped dynamics' of the points in ${\cal R}$
driven by the following potential function
\[
\Delta \equiv \sum(T_n -n)^2. \label{Dd}
\]
Here, the summation is over all the pairs. In the actual
implementation of the algorithm, simpler forces are adopted than
the one obtained from this potential as seen below, because the
latter is  complicated for numerical simulation to the extent of
being not practical.  There are many ways to modify $\Delta$ to
make the force simpler; we have selected the one that makes the
force formula very simple, if not the simplest.

The $\Delta$ defined by the formula above may be regarded as a
counterpart of the stress in the conventional nMDS. As we will
see later we can use quantities related to $\Delta$ to evaluate
the confidence level of the resultant configuration.   However, we
do not employ $\Delta$ itself, because this quantity is not the difference
between two ranks of independent objects; there are $_{N}C_2$
$\delta_{ij}$ or $d_{ij}$ but they are relations among only $N$
potentially independent quantities.  Therefore. the statistics of
$\Delta$ is not known.

Thus, although certain modifications as discussed above are
needed, still in our nMDS algorithm, the optimization process is
closely connected to a process that improves the confidence level
of the resultant configuration.

The `pure nMDS' algorithm for $N$ objects may be described as follows:
\begin{enumerate}
\item Dissimilarities $\delta_{ij}$ $(i,j = 1, \cdots, N)$
          for $N$ objects are given. Order them as follows:
\[
          \cdots \leq \delta_{ij} \leq \delta_{kl} \leq \cdots.\label{delta}
\]
\item Put $N$ points randomly in ${\cal R}$ as an
          initial configuration.
\item Scale the position vectors in ${\cal R}$ such as
$\sqrt{\sum_i {|\vec r}_i|^2}=1$,    where ${\vec r}_i$ is the
current position of object $i$ in ${\cal R}$.
\item Compute $d_{ij}$ for all object pairs $(i,j)$
          in ${\cal R}$, and then order them as
\[
          \cdots \leq d_{ij} \leq d_{kl} \leq \cdots.\label{do}
\]
\item Suppose $\delta_{ij}$ is the $m$th largest in
          the ordering in \ref{delta} and $d_{ij}$ is the $T_m$th
          largest in the ordering  in \ref{do}. Assign $C_{ij}= T_m-m$.
Calculate the following displacement
          vector for $i$:
\[
          \d\vec{r}_i = s \sum_j C_{ij} \frac{{\vec r}_i-{\vec r}_j}{|{\vec
r}_i-{\vec r}_j|},
\]
where $s=0.1 \times N^{-3}$ typically, and update $\vec{r}_i \map
\vec{r}_i + \d \vec{r}_i$.
\item Return to 3, and continue until the ``potential energy
$\Delta$'' becomes sufficiently small.
\end{enumerate}

The reader may worry about the handling of  tie data.
Generally speaking, for a large data set the fraction of tie
relations is not significant; furthermore, if the result depends on
the handling schemes of tie data, the result is unreliable anyway.
Therefore, we do not pay any particular attention to the tie data
problem.

In the above algorithm, $s$ is a constant value.  In practice, we
could choose an appropriate schedule to vary $s$ as is often done
in optimization processes.  In this paper, for simplicity, we do not
attempt such a fine tuning.

In the above algorithm, we can deal with asymmetric data as
well, i.e., $\delta_{ij} \neq \delta_{ji}$ if we compare
$\delta_{ij}$ with $d_{ij}$ while $\delta_{ji}$ with $d_{ji}
(=d_{ij})$.  Needless to say, if the mismatch between
$\delta_{ij}$ and $\delta_{ji}$  is large, then representing the
pair by a pair of points in a metric space is questionable.
Therefore, we will not discuss this problem any further in this paper.
The microarray data analyzed in this paper do not have such a
problem in any case.

\noi
{\bf Goodness of embedding}\\
In the text we have already discussed the effective number of
correctly embedded objects as a measure of `global goodness of
embedding.'   This measure, however, cannot tell us the embedding
quality  of each object.  It is often the case that the majority of
objects are embedded well even without sensitive dependence on
the initial conditions, but there are a few objects that
consistently refuse to be embedded stably.  To judge the
quality of embedding for each object $j$ we define
\[
\Delta(j) \equiv \sum \left[T_{n(j)}(j) -n(j)\right]^2,  \label{D}
\]
Here, $n(j)$ is the rank order of $\d(i,j)$ among $N-1$
pairs $(i,j)$ for a given  $j$, and $T_{n(j)}(j)$ is the rank order of
$d(i,j)$ among $N-1$ pairs $(i,j)$ for the same $j$.

$\Delta(j)$ can be regarded as a statistical variable for the
relative position of the $j$-th object with respect to the
remaining objects (Lehmann 1975).  We can estimate the
probability $P(\e)$ of $\Delta(j)<\e$ with the null hypothesis that
the rank ordering of $d_{ij}$ ($ i \in \{1,2, \cdots, N\}\setminus
\{j\}$) is totally random with respect to the rank ordering of
$\delta_{ij}$ ($ i \in \{1,2, \cdots, N\}\setminus \{j\}$). If $N$
is sufficiently large, then $\Delta(j)$ obeys the  normal
distribution with mean $(M^3-M)/6$ and variance
$M^2(M+1)^2(M-1)/36$, where $M \equiv N-1$.  For smaller $N$
there is a table for $P(\epsilon)$ (Lehmann 1975).  Thus, we can
test the null hypothesis with a given confidence level for $j$-th
object.

\noi
{\bf APPENDIX II}\\
{\bf Limitations and capabilities of linear methods}\\
The limitations and capabilities of PCA with and without data
preprocessing are illustrated in this appendix.  There is no
fundamental difference between PCA and SVD. We consider the
following artificial data $\{s_{gt}\}$, where $g$
($=1,\cdots,517$) denote genes and $t$ ($=1, \cdots, 11$) the
observation times:\\
$ $[Data set 1]
\[
s^1_{gt} = C_g \cos (2 \pi t / 11 + 2 \pi \delta_g).
\]
$ $[Data set 2]
\[
s^2_{gt} = \exp(s^1_{gt}).
\]
$ $[Data set 3]
\begin{eqnarray*}
s^3_{gt} &= &C_{g1} \cos (2 \pi t / 11 + 2 \pi \delta_{1g}) \\
& & + C_{g2} \exp[\cos (2 \pi t / 11 + 2 \pi \delta_{2g})] \\
& & \mbox{}\mbox{} + C_{g3}/ \cos (2 \pi t / 11 + 2 \pi \delta_{3g}).
\end{eqnarray*}
In the above, $C_g, \delta_g, C_{ig},\delta_{ig},(i=1,2,3)$ are
uniform random numbers in $[0,1]$.  That is, Data set 1 is a set of
sinusoidal waves with random amplitudes and phases, Data set 2
is the nonlinearly distorted Data set 1, and Data set 3 is a set of
periodic functions that are very different from simple oscillatory
behaviors.

These data sets are analyzed by the following methods.\\
\noi
Method 1: PCA with the preprocessing used by Holter {\it et al}.
(2000). The preprocessing procedure is as follows:\\
step 1: Subtract the average,
$$
s'_{gt}= s_{gt} - \langle s_{gt}\rangle_{g,t},
$$
where  $\langle \bullet \rangle_{g,t}$ is the average over all genes and
experiments,
$$
\langle \bullet \rangle_{g,t} \equiv  \frac{\sum_{g,t} \bullet}{\sum_{g,t}1}.
$$
\\
step 2: (Column normalization) Normalize the data as
$$
s''_{gt} = \frac{s'_{gt}}{\sqrt{\left[\sum_g (s'_{gt})^{2}\right]}} .
$$\\
step 3: (Row normalization) Normalize the data as
$$
s'''_{gt} = \frac{s''_{gt}}{\sqrt{\left[\sum_t ( s''_{gt})^{2}\right]}} .
$$\\
Repeat these steps until the following condition is satisfied,
$$
\sqrt{\langle \left[s_{gt}-s'''_{gt}\right]^2 \rangle_{g,t}} < 0.01.
$$
\\
       From the resultant $s_{gt}$ correlation matrix
$Matr.(Cor_{t t'})$
is constructed, and then PCA is performed.

\noi
Method 2: PCA with the preprocessing
so that $\sum_t s_{gt} =0 $ and $\sum_t s_{gt}^2=1$ for all $g$.
Of course, no iteration is needed for this preprocessing.  From the
resultant $s_{gt}$ correlation matrix $Matr.(Cor_{t t'})$  is
constructed, and then PCA is performed.

\noi
Method 3: nMDS as done in the text.  That is, the negative of the
correlation coefficient $Cor_{g g'}$ is used as the dissimilarity
and nMDS is applied straightforwardly. Needless to say, no
preprocessing of data is needed.

\marginparsep -9mm
\marginpar{\fbox{Fig. 7}}
\marginparsep 4mm

The results are exhibited in Figure 7.
   The conclusions may be:\\
(1) For Data set 1, any method will do.\\
(2) For Data set 2, the procedure recommended by Holter {\it et
al.} (2000) fails, although ironically simpler Method 2 still works
very well.  If the amplitude $C$ is distributed in $[0,5]$ instead
of $[0,1]$ (that is, the extent of the nonlinear distortion is
increased), Method 2 becomes inferior to Method 3, but still
Method 2 is adequate.\\
(3) For Data set 3, even Method 2 fails.
nMDS (Method 3) still exhibits a ring-like structure. The method
recommended by Holter {\it et al}.\ (2000) is obviously out of question.

Thus, we may conclude that nMDS is a versatile and all around
data mining method for analyzing periodic temporal data.
Furthermore, we can point out that the preprocessing method in
Method 1 should not be used,
because it could severely distort the
original data (as may have been expected from the figures).
Suppose there are $N$ genes and $4$ time points.  Consider the
following example (for the counterexample sake). The first gene
has $(a,b,-b,-a)$ ($a>b>0$), and the remaining genes are all give
by $(1, 0,0,-1)$.  The $N \times 4$ matrix made from these
vectors is polished by an iterative row and column vector
normalization procedure.  If $N$ is sufficiently large, the
first row converges to $(0,1,-1,0)$ and the rest to $(1,0,0,-1)$,
independent of $a$ and $b$.  If $b$ is small, then all the vectors
should behave almost the same way, but after polishing the
out-of-phase component in the discrepancy between the first row and
the rest is dramatically enhanced, resulting in a spurious out of
phase temporal behavior.  Although the preceding exercise is
trivial, the result  warns us the danger of using the so-called
polishing.

\noi
{\bf APPENDIX III}\\
\noi
{\bf  Gene ontology}\\
Since we have seen Laegreid
{\it et al}.\
(2003)
claiming that gene ontology can be discovered in the same data we
are analyzing, we tried
to extract more biological
information from the nMDS result.  We have realized that the
distribution of ontologically identified genes (or that of the
distribution of 23 ontology tags adopted by Laegreid
{\it et al.}) along
the ring is statistically indistinguishable from the uniform
distribution.  In this sense, nMDS fails to capture ontological
information of the genes.  This might suggest that our nMDS
method is definitely inferior to the analysis based on the rough
set adopted by the above paper. However, we have found that the
claimed success is largely illusory.  The claimed success is: of
the 24 genes for which homology information allows inference of
their ontology, ``11 genes had one or more classifications that
matched this assumption (Table 10).''  However,  actually 8 $\pm$
2.4 is within one $\sigma$ error (note that out of 210 genes used
to train the rules based on the rough set, each gene on the average
share an ontology with about 70 other genes).  That is, 11 out of
24 is statistically hardly significant (the cross validation does
not mean very much because it is only a consistency check of the
method).  Therefore, nMDS does not fail to capture the significant
information that can be obtained by the existing statistical methods.

Thus, we may conclude that the data we have analyzed may be
sufficiently informative to follow the cell cycle related events,
but not to annotate genes ontologically.

\noi
{\bf REFERENCES}\\
\noi
Borg, I.,  and Groenen, P.  (1997) {\it Modern Multidimensional
        Scaling}, Springer, New York.

\noi
        Cox, T. F., and Cox,  M. A. A., (1994) {\it Multidimensional Scaling},
Chapman \& Hall, London.

\noi
Cho,  R.J., Huang, M., Campbell,  M. J., Dong,  H., Steinmetz,  L.,
Sapinoso. L., Hampton, G., Elledge,  S. J.,
Davis, R. W., and Lockhart, D., J., (2001)
Transcriptional regulation and function during the human cell cycle,
{\it Nat. Genet.}, {\bf 27}, 48-54.

\noi
Donoho, D.L.,  Vetterli,M.,  DeVore,R.A.,  and Daubechies, I. (1998)
Data Compression and Harmonic Analysis,
{\it IEEE Transactions on Information Theory}, {\bf 44},
2435-2476.

\noi
Dyrskjot, L., Thykjaer, T., Kruhoffer. M.,
Jensen,  J. L., Marcussen, N., Hamilton-Dutoit,
S, Wolf, H., and Orntoft, T. F.,
   (2002), Identifying distinct classes of
bladder carcinoma using microarrays,
{\it  Nature genetics}, {\bf  33},  90 - 96.

\noi
Eisen, M. B., Spellman, P.T., Brown, P. O., and Botstein D., (1998)
           Cluster analysis and display of genome-wide expression patterns,
{\it Proc. Natl. Acd. Sci. USA}, {\bf 95}, 14863-14868.

\noi
Green, P.E.,  Carmone Jr., F. J.,   and Smith, S. M., (1970)
{\it Multidimensional Scaling : Concepts and Applications},
Allyn and Bacon, Massachusetts.

\noi
Hollander, M., and Wolfe, D. A., (1999)
{\it Nonparametric Statistical Methods},
        John Wiley \& Sons, New York.

\noi Holter, N. S., Mitra, M., Maritan, A., Cieplak, M., Banavar,
J. R., and Fedoroff, N. V.,  (2000)
        Fundamental patterns underlying gene expression profiles: Simplicity
from complexity,
{\it Proc. Natl. Acad. Sci. USA},  {\bf 97}, 8409-8414.

\noi
Iyer,  V. R.,   Eisen, M. B.,   Ross,  D. T.,
        Schuler,  G.,  Moore, T., L.,   Jeffrey C. F.   Trent,  J. M.,
Staudt, L. M.,  Hudson Jr.,   J.,   Boguski,   M. S.,
Lashkari, D., Shalon, D.,   Botstein,  D.,  and  Brown,    P. O.,  (1999)
The Transcriptional Program in the Response of Human Fibroblasts to Serum,
{\it Science}, {\bf 283}, 83-87.

\noi
Johansson, D.,  Lindgren, P.,  and Beglund, A.,  (2003)
A multivariate approach applied to microarray data for
identification of genes with cell cycle-coupled transcription,
{\em Bioinformatics}, {\bf 19}, 467-473.

\noi
Kanaya, S., Kinouchi M., Abe, T., Kudo, Y., Yamada, Y., Nishi, T., Mori,
Mori, H. and Ikemura, T., (2001) Analysis of codon usage diversity of
bacterial genes with a self-organizing map (SOM): characterization of
horizontally transferred genes with emphasis on the E. coli O157 genome,
{\it Gene}, {\bf 276}, 89-99.

\noi
Kasturi, J., Acharya, R.,  and Ramanathan, M., (2003)
An information theoretic approach for analyzing temporal patterns of gene
expression, {\it Bioinformatics}, {\bf 19},  449-458.

\noi Kruskal, J. B., (1964a) Multidimensional scaling by optimizing goodness of
fit to a nonmetric hypothesis, {\it Psychometrika}, {\bf 29}, 1-27.

\noi Kruskal, J. B., (1964b)
Nonmetric multidimensional scaling: A numerical method, {\it Psychometrika},
{\bf 29}, 115-129.

\noi
Leagreid, A.,  Hvidsten, T. R.,
Midelfart, H., Komorowski, J.,  Sandvik, A. K.,
(2003) Predicting gene ontology biological process
from temporal gene expression patterns,
{\it Gen. Res.} {\bf 13}, 965-979.

\noi
Lehmann, E. L., (1975) {\it Nonparametrics}, Holden-Day, San Francisco.

\noi Shedden, K. and Cooper, S. (2002)
Analysis of cell-cycle-specific gene expression in
human cells as determined by microarray and double-thymidine block
synchronization, {\it Proc. Natl. Acad. Sci. USA}, {\bf 99}, 4379-4384.

\noi Shepard, R. N., (1962a)
The analysis proximities: Multidimensional scaling
with an unknown distance function, I {\it Psychometrika}, {\bf 27}, 125-140.

\noi Shepard, R. N., (1962b)
The analysis proximities: Multidimensional scaling
with an unknown distance function, II {\it Psychometrika}, {\bf 27}, 219-246.

\noi Shmulevich, I. and Zhang, W, (2002) Binary analysis and
optimization-based normalization of gene expression data,
{\it Bioinformatics}, {\bf 18}, 555-565.

\noi
Slonim, D. K., (2002) From patterns to pathways: gene
expression data analysis of age, {\it Nature Genetics
Supplement}, {\bf 32}, 502-508.

\noi
Spellman, P. T.,  Sherlock, G., Zhang, M. Q.,
Iyer, V. R., Anders, K.,  Eisen, M. B.,
Brown, P. O., Botstein, D., and Futcher, B.,
(1998)
        Comprehensive Identification of Cell Cycle regulated Genes of the Yeast
{\it Saccharomyces cerevisiae}
by Microarray Hybridization,
{\it Molecular Biology of the Cell}, {\bf 9},  3273-3297.

\noi
Taguchi, Y-h., and  Oono, Y., (1999), unpublished.
\\http://www.granular.com/MDS/src/paper.pdf

\noi
Taguchi, Y-h., and  Oono, Y., (2004),
Adv. Chem. Phys. in press.

\noi
Taguchi, Y-h., Oono, Y., and Yokoyama, K., (2001)
New possibilities of non-metric multidimensional scaling,
{\it Proc. Inst. Stat. Math.}, {\bf 49}, 133-153 (in Japanese).

\pagebreak
\onecolumn

Table legends

Table 1:
The dependence of the effective number $N'$ of correctly embedded
points upon the number $M$ of geometrically embeddable points.  The
$N'=100$  case may reflect the fact that $N'$ is a good estimate only
for sufficiently many embeddable points.

Table 2:
Relationship between gene functions and the two-dimensional arrangement.

Table 3:
The correlation coefficients between principal axes of embedding
in  $D$-dimensional space. For detail, see text.

\pagebreak

\noindent
Figure legends

\noindent
Figure 1:
PCA results using correlation coefficient matrix.
The first two principal components are used as the horizontal and
vertical axis, respectively (the cumulative proportion is 70 \%).
Genes whose experimental values are larger than $3.2$ are drawn
using filled boxes, otherwise, using small dots
(the corresponding color figure is available online).
       From the top the time is, respectively, 15 min, 30 min, 1 hr, 2 hr,
4 hr, 6 hr, 8 hr, 12 hr, 16 hr, 20 hr, 24 hr.  The figures are
arranged in two columns, but this is solely for the layout purpose;
there is no distinction between two columns.

\noindent
Figure 2:
Two dimensional embedding  result obtained by nMDS.

\noindent
Figure 3:
Comparison between the nMDS embedding results with
two different initial configurations after Procrustean similarity
transformation.
The horizontal (resp., vertical) coordinates
are compared in the  left (resp., right) figure.
In each figure $x$-axis corresponds to the result from
one initial condition and the $y$-axis the other.

\noindent
Figure 4:
(a) Temporal patterns of gene expression levels visualized with the aid of
nMDS.
Genes whose experimental values are larger than $1.2$ are drawn
using filled boxes, otherwise, using small dots
Time sequences are the same as explained at Fig.\ 1.
\noindent
(b) Gene expression data as a function of the angle measured from the
vertical axis in (a).
The
horizontal axis corresponds to $t$.
Genes whose experimental values are larger than $1.6$ are drawn
using filled boxes, otherwise, using small dots
The figures are arranged in two columns, but this
is solely for the layout purpose; there is no distinction between
two columns.

\noindent
Figure 5:
3D unfolding of the temporal pattern of gene expression level with the aid of
nMDS (3D).
Experimental values are normalized as explained in Fig.\ 3.
Genes whose experimental values are larger than $1.6$ are drawn
using filled boxes, otherwise drawn using small dots
(the corresponding color figure is available online).
The horizontal (resp., vertical) axis represents $\phi$ (resp., $\theta$).
See the text for detail.

\noindent
Figure 6:
Example of less well structured data.
Upper row: nMDS;
Middle row PCA with normalization (Method II in Appendix II);
Lower row:
Most expressive/depressive genes (with larger variance):
Left column: original data;
Right column: randomized data.
For detail, see text.

\noindent
Figure 7:
Comparison of linear and nonlinear methods.\\
Method 1: PCA with polishing (Holter {\it et al}.\ 2000);
Method 2: PCA with normalization;
Method 3: 2$D$ space embedding with the aid of nMDS.
See the text for Data sets and Methods.
For Methods 1 and 2, horizontal and vertical axes
are the first and second principal components, respectively, and
the percentages describe cumulative proportions.
For Method 3, the percentages are the indicators of goodness defined in the
text.

\pagebreak

Table 1 Taguchi \& Oono

\begin{tabular}{cccc}
$M$  &    $N'$ &     $N'-M$ &    $\sqrt{N-M}$ \\\hline
180 & 183.6  &  +3.6 &  4.5 \\
170 & 174.8  &  +4.8 &  5.5 \\
150 & 158.4 &  +8.4 &  7.1 \\
120 & 127.9 & +7.8 & 8.9 \\
100 & 89.6  & -10.4 & 10.0
\end{tabular}

\pagebreak

\vspace{-3cm}

Table 2 Taguchi \& Oono

\begin{tabular}{l|l|l|l|l}
\hline
Angle & \begin{minipage}{2cm}
Gene Number in   Iyer
{\it et al}.\ (1999)\end{minipage} &
Gene identification & cell cycle & notes
\cr
\hline
1.83 & 332&	RRM1 &	S &
\begin{minipage}[t]{7cm}
DNA synthesis
\end{minipage}
\cr
1.84&342 &	Cyclin B1  &
G2/M &
\begin{minipage}[t]{7cm}
predominantly during G2/M phase
\end{minipage}
\cr
1.85&331 &	centromere protein F	& G2-M&
\begin{minipage}[t]{7cm}
in late G2  through
early anaphase; the product associates with the kinetochore.
\end{minipage}
\cr
1.90 &339 &	CDC2 & G2/M &
\begin{minipage}[t]{7cm}
essential for G1/S and G2/M transitions
controlled by cyclin accumulation and destruction through the cell
cycle.
\end{minipage}
\cr
1.91& 328&	MAD2L1& M &
\begin{minipage}[t]{7cm}
a component of the mitotic spindle
assembly checkpoint that prevents the onset of anaphase until all
chromosomes are properly aligned at the metaphase plate.
\end{minipage}
\cr
1.92&333 &
CCNA2& G2/M &
\begin{minipage}[t]{7cm}
transition G1/S and G2/M. This cyclin
binds and activates CDC2 or CDK2 kinases, and thus promotes both cell
cycle G1/S and G2/M transitions.
\end{minipage}
\cr
1.93&329 &	yanuru& M &
\begin{minipage}[t]{7cm}
encoding proliferation-associated nuclear
protein; associates with the spindle pole and mitotic spindle during
mitosis.
\end{minipage}
\cr
1.95 & 341&
CKS2 &
G2/M&
\begin{minipage}[t]{7cm}
encoding CDC28 protein kinase regulatory
subunit 2. CKS2 protein binds to the catalytic subunit of the cyclin
dependent kinases and is essential for their biological function.
\end{minipage}
\cr
2.10 &308 &	CENPE& M &
\begin{minipage}[t]{7cm}
encoding centromere protein E, 312kDa;
first appears in prometaphase of M.
\end{minipage}
\cr
2.51 &306 &	LBR& M &
\begin{minipage}[t]{7cm}
encoding lamin b receptor; chromatin organization.
\end{minipage}
\cr
2.64&204 &	CDKN2C& G1&
\begin{minipage}[t]{7cm}
encoding cyclin-dependent kinase
inhibitor 2C (p18, inhibits CDK4); controls G1.
\end{minipage}
\cr
2.68& 319 &	CNAP1&M &
\begin{minipage}[t]{7cm}
encoding chromosome condensation-related
condensin homolog.
\end{minipage} \\
\hline
\end{tabular}
\pagebreak


Table 3 Taguchi \& Oono

\begin{center}
\begin{tabular}{l|ccc}
\hline
&3DI & 3DII & 3DIII \\
        \hline
        2DI & 0.998 & 0.014 & 0.005\\
        2DII & $-$0.014 & 0.993 & $-$0.027\\
        \hline
        \end{tabular}\\
\end{center}

\pagebreak
\hfill Figure 1 Taguchi \& Oono

\vspace{-2cm}
\begin{tabular} {rccl}

\raisebox{1.5cm}{15 min.}&
\includegraphics[width=3cm]{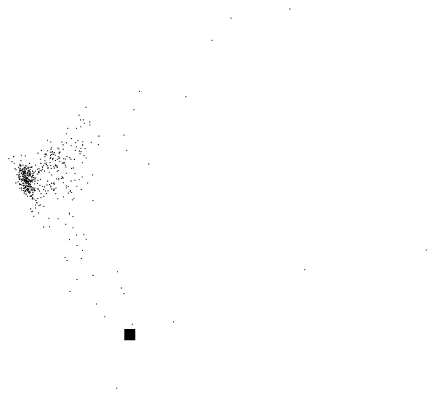} &&\\

\raisebox{1.5cm}{30 min.}& \includegraphics[width=3cm]{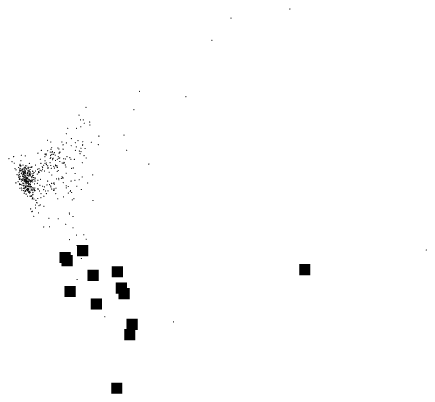} &
\includegraphics[width=3cm]{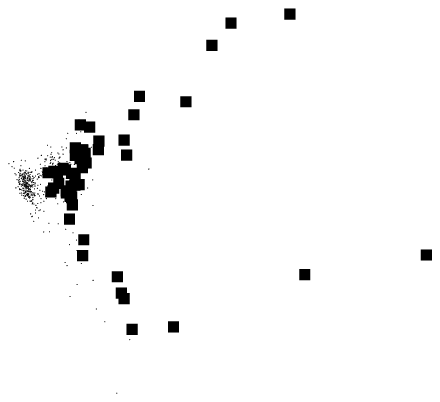} & \raisebox{1.5cm}{6 hr}\\

\raisebox{1.5cm}{1 hr}& \includegraphics[width=3cm]{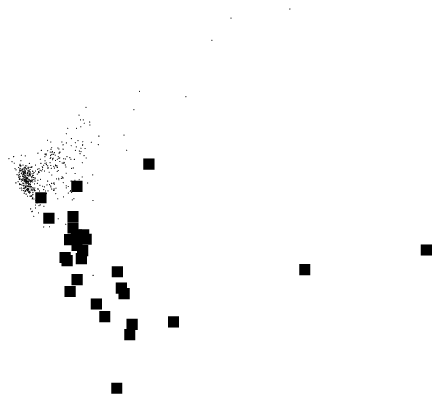} &
\includegraphics[width=3cm]{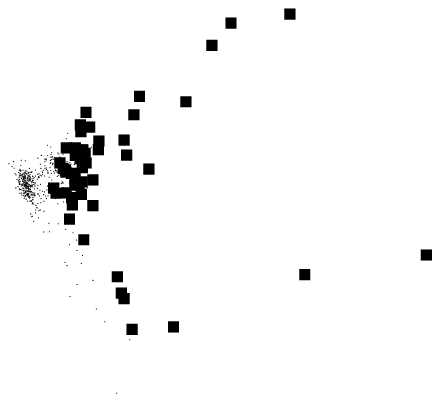}& \raisebox{1.5cm}{8 hr} \\

\raisebox{1.5cm}{2 hr}& \includegraphics[width=3cm]{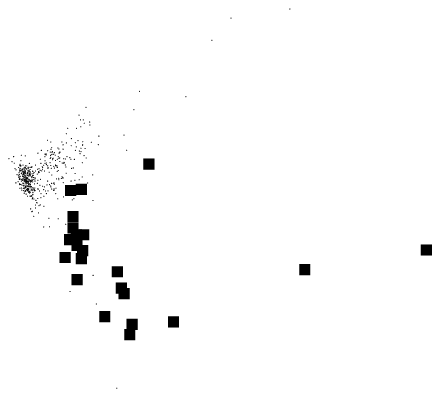} &
\includegraphics[width=3cm]{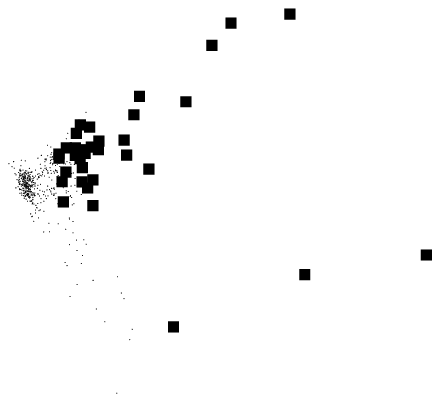}& \raisebox{1.5cm}{12 hr}\\

\raisebox{1.5cm}{4 hr}& \includegraphics[width=3cm]{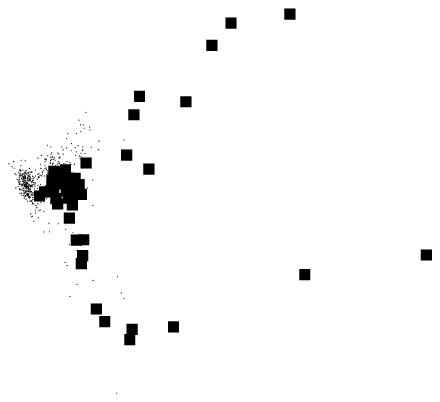} &
\includegraphics[width=3cm]{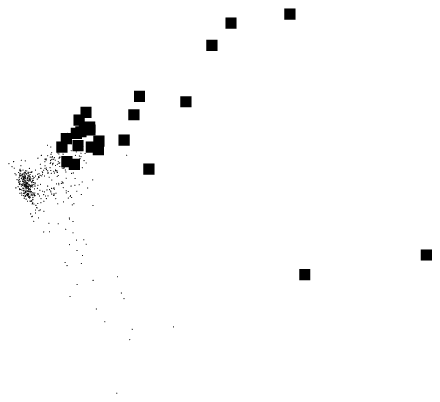} & \raisebox{1.5cm}{16 hr} \\





& & \includegraphics[width=3cm]{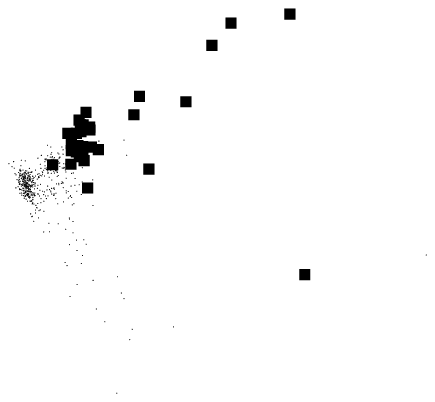}&
\raisebox{1.5cm}{20 hr} \\

&&\includegraphics[width=3cm]{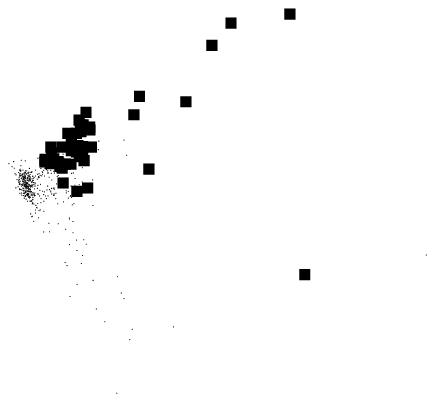}& \raisebox{1.5cm}{24 hr} \\
\end{tabular}

\pagebreak

Figure 2.  Taguchi \& Oono

\bigskip

\begin{tabular}{cc}
\rotatebox{90}{\hspace{2cm}$y$} &\includegraphics[width=5cm]{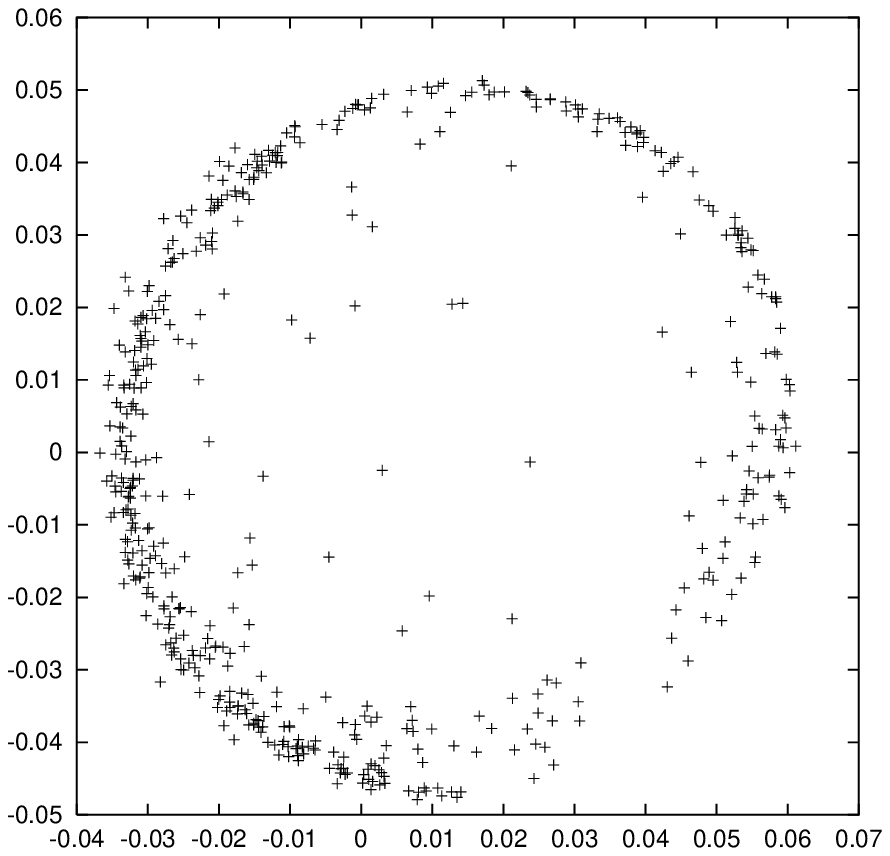} \\
& $x$
\end{tabular}

\pagebreak

Figure 3  Taguchi \& Oono

\bigskip

\rotatebox{90}{coordinate (one initial condition)}
\includegraphics[width=7cm]{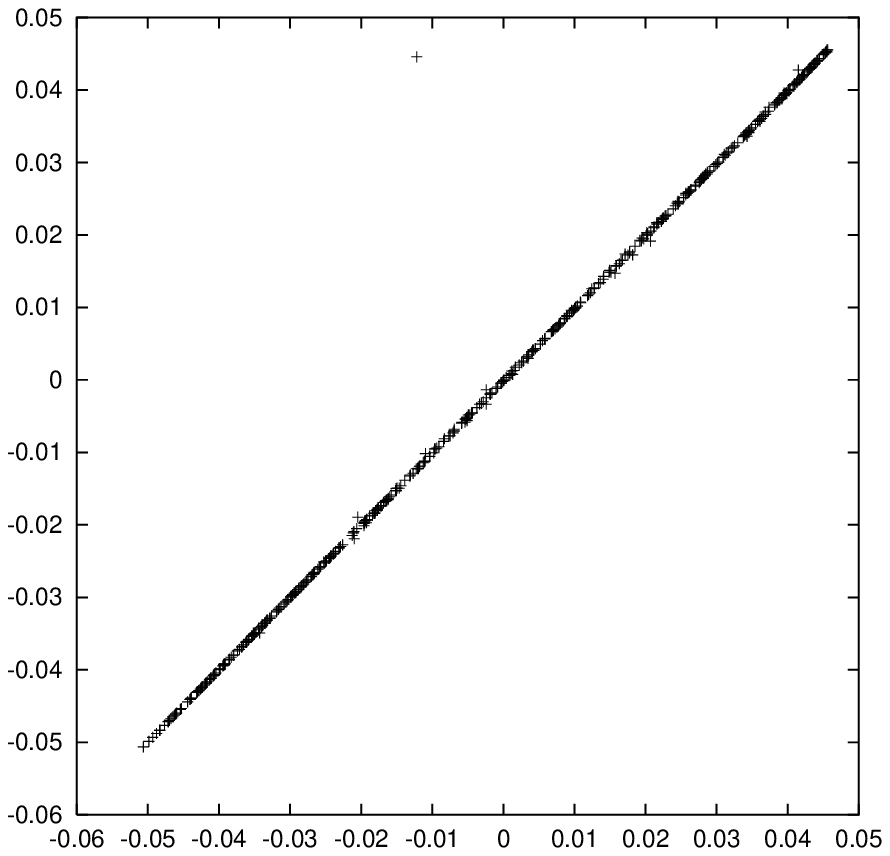}
\includegraphics[width=7cm]{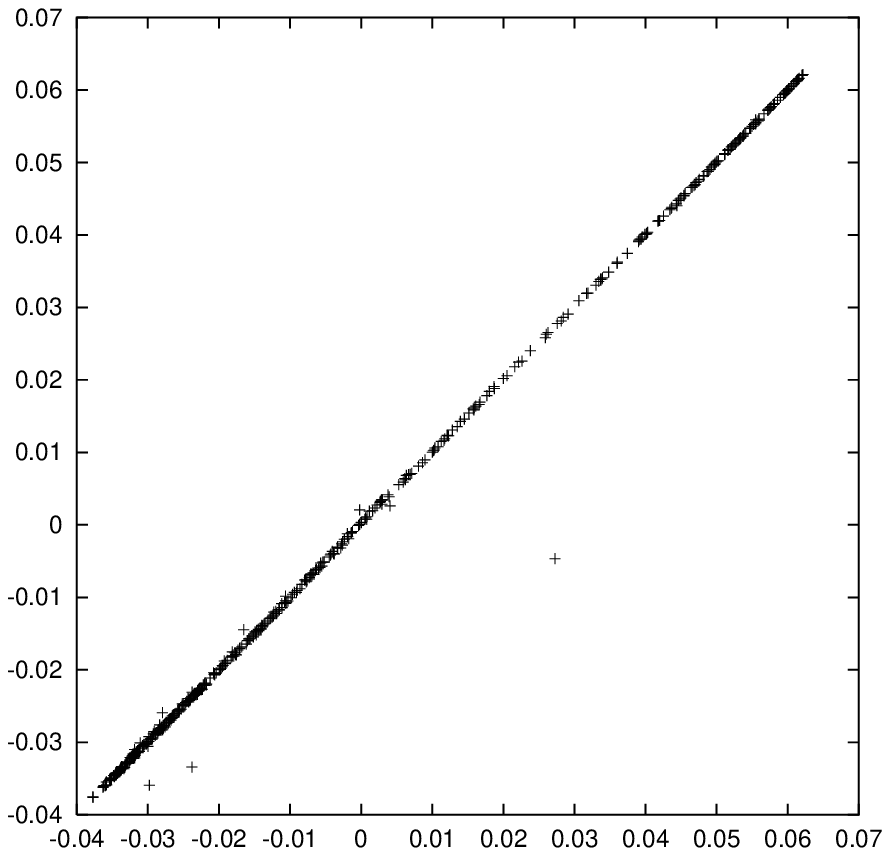}

\hspace{5cm}coordinate (another  initial condition)

\pagebreak


\begin{tabular} {rccl}

\raisebox{3cm}{(a)}
\raisebox{1.5cm}{15 min.}& 
\includegraphics[width=3cm]{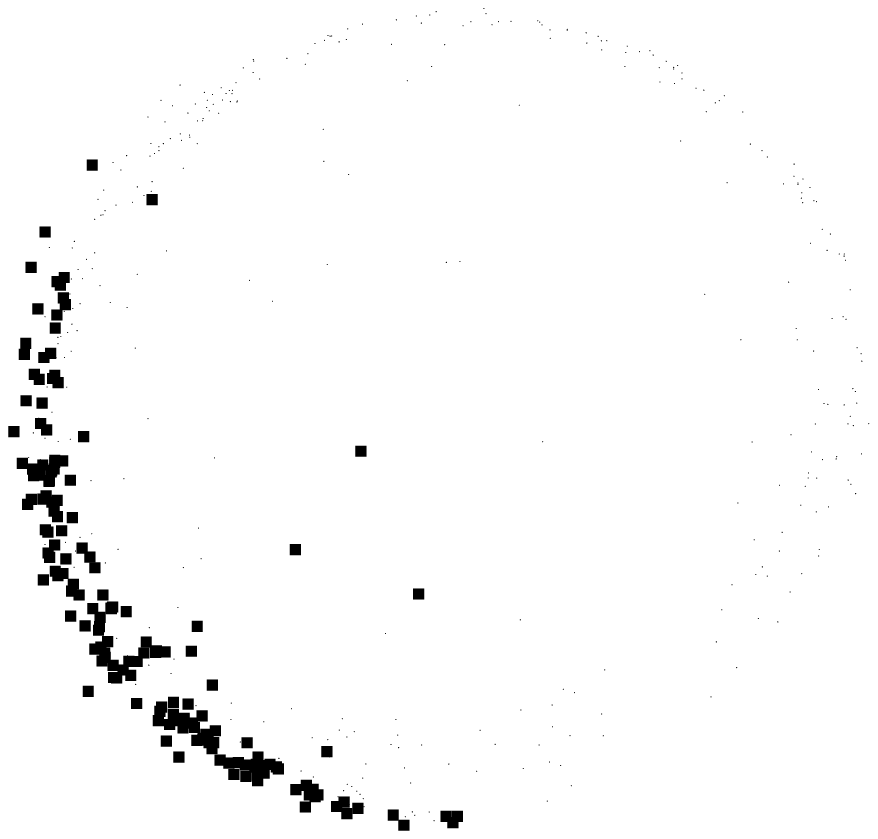} &&\\

\raisebox{1.5cm}{30 min.}& \includegraphics[width=3cm]{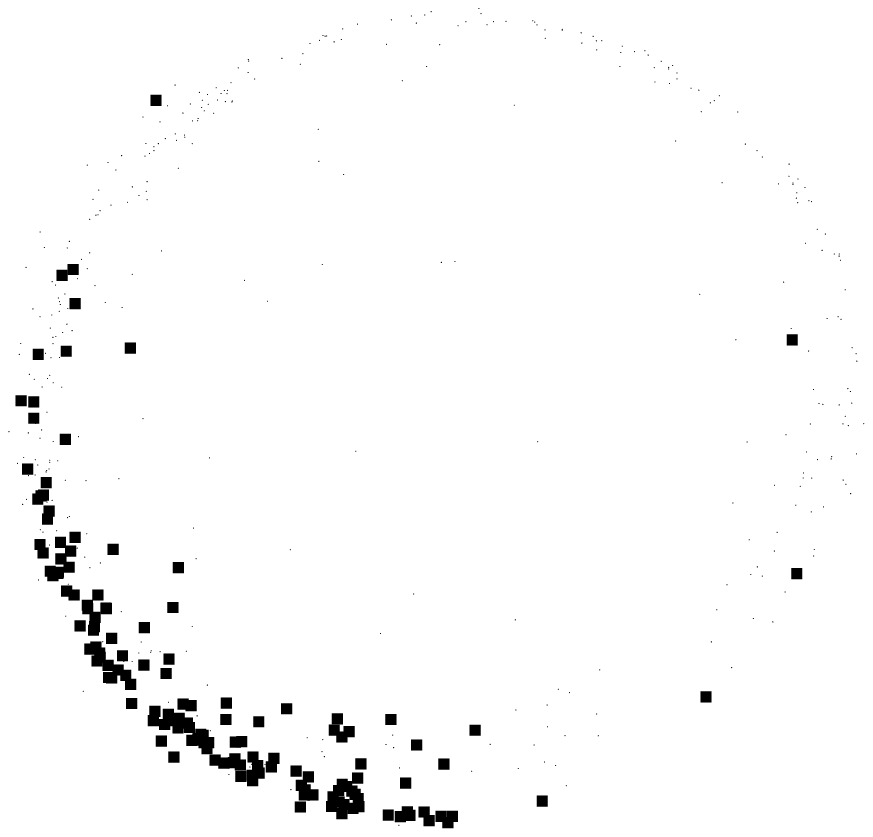} &
\includegraphics[width=3cm]{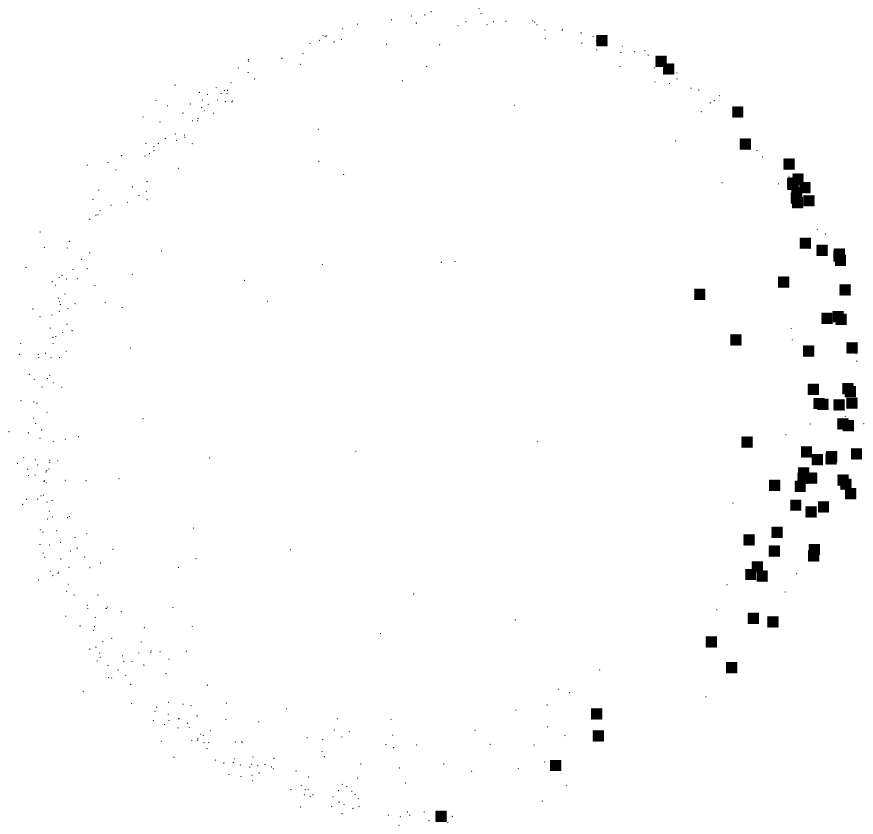} & \raisebox{1.5cm}{6 hr}\\

\raisebox{1.5cm}{1 hr}& \includegraphics[width=3cm]{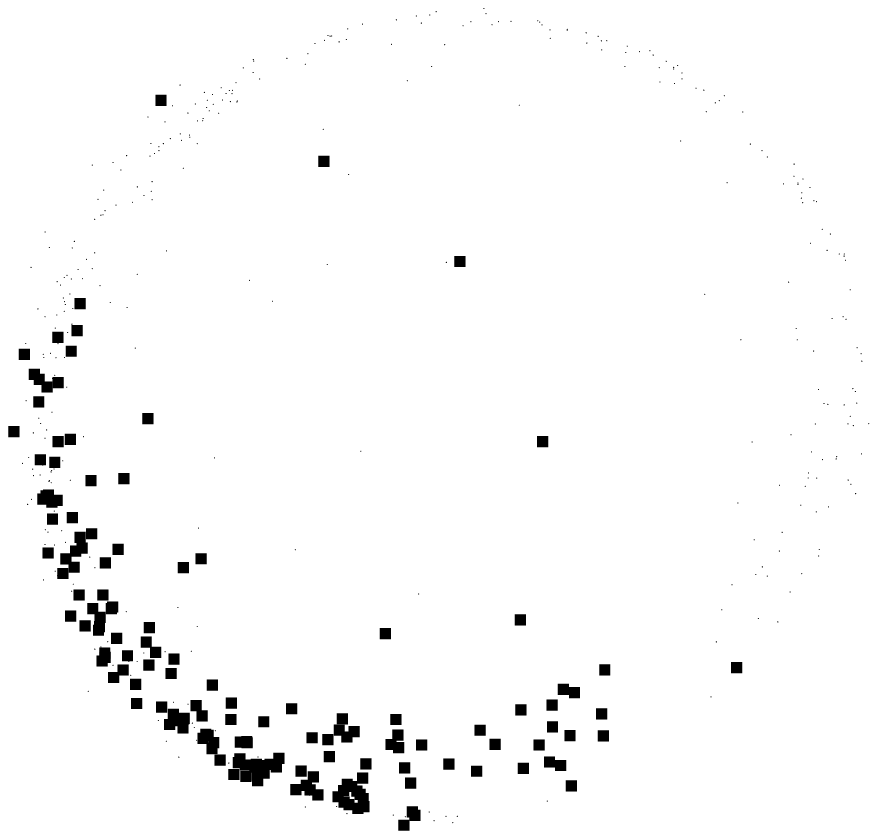} &
\includegraphics[width=3cm]{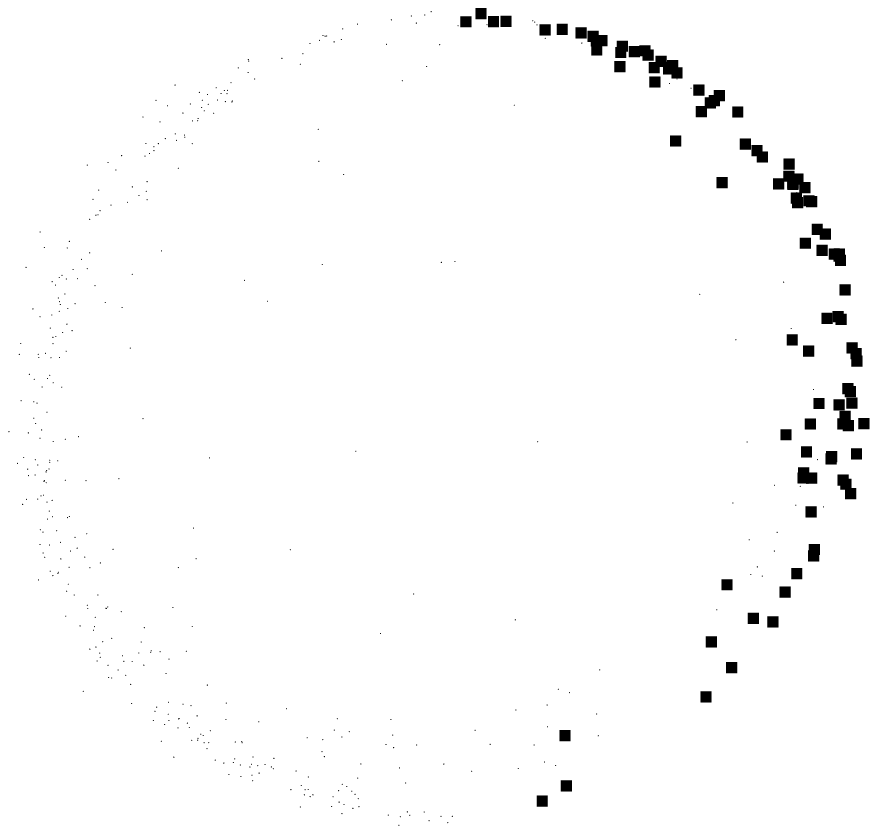}& \raisebox{1.5cm}{8 hr} \\

\raisebox{1.5cm}{2 hr}& \includegraphics[width=3cm]{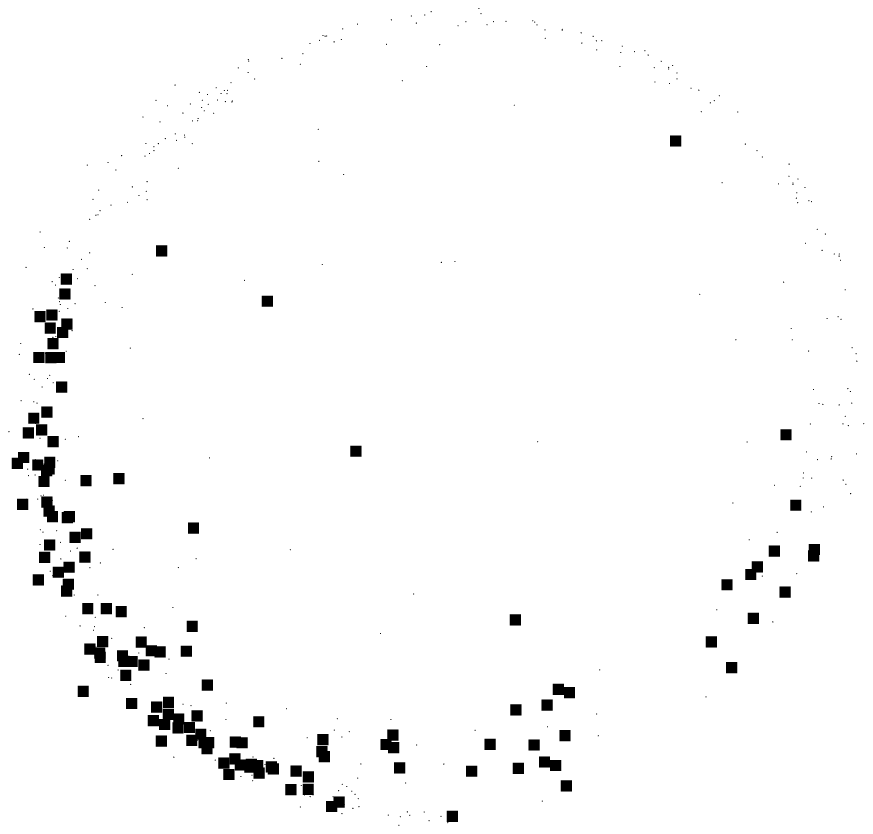} &
\includegraphics[width=3cm]{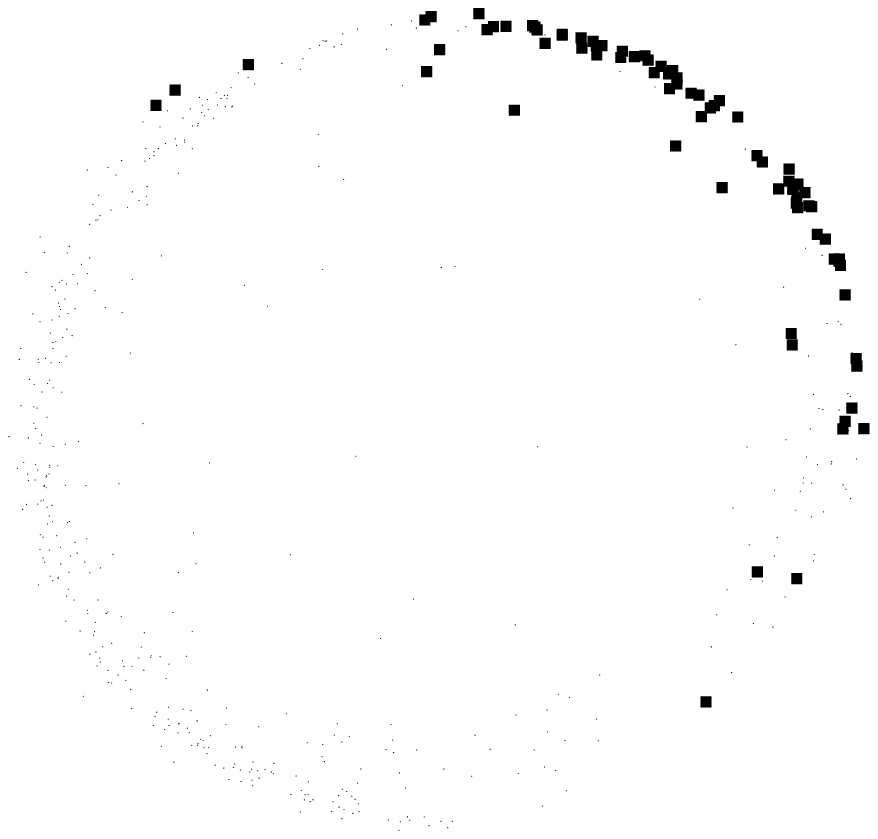}&\raisebox{1.5cm}{12 hr}\\

\raisebox{1.5cm}{4 hr}& \includegraphics[width=3cm]{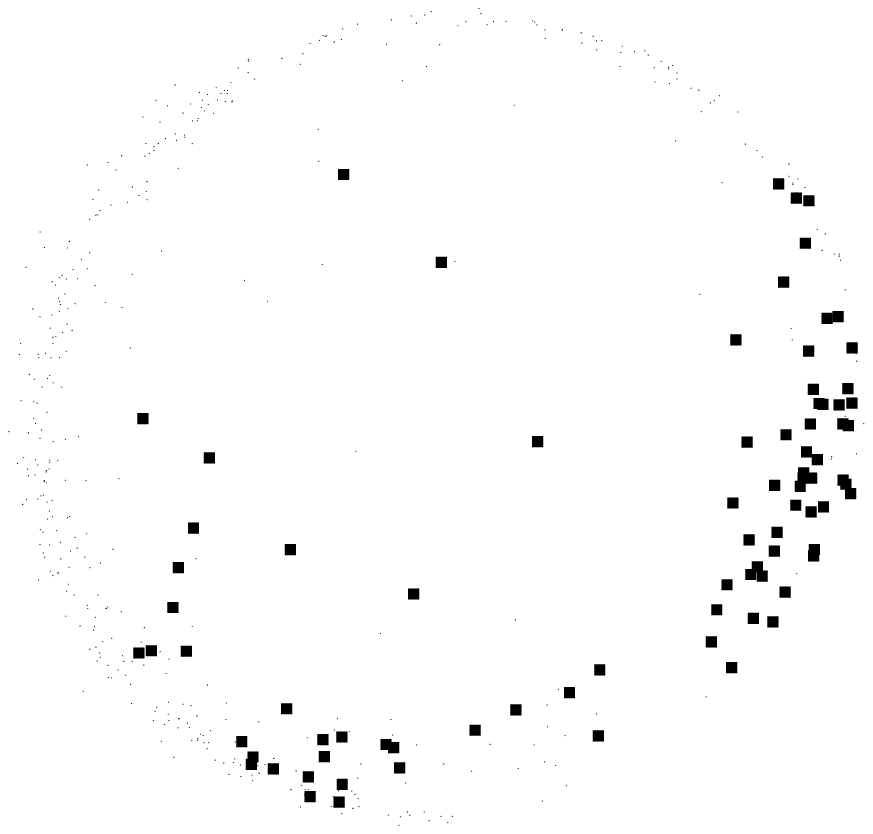} &
\includegraphics[width=3cm]{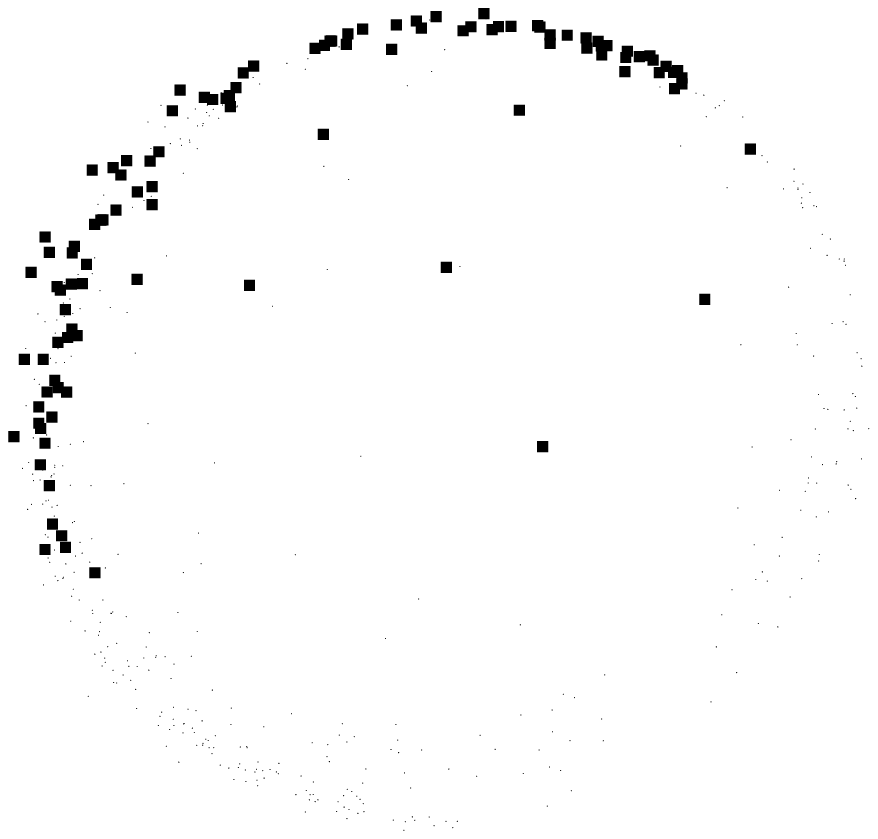} & \raisebox{1.5cm}{16 hr} \\





& & \includegraphics[width=3cm]{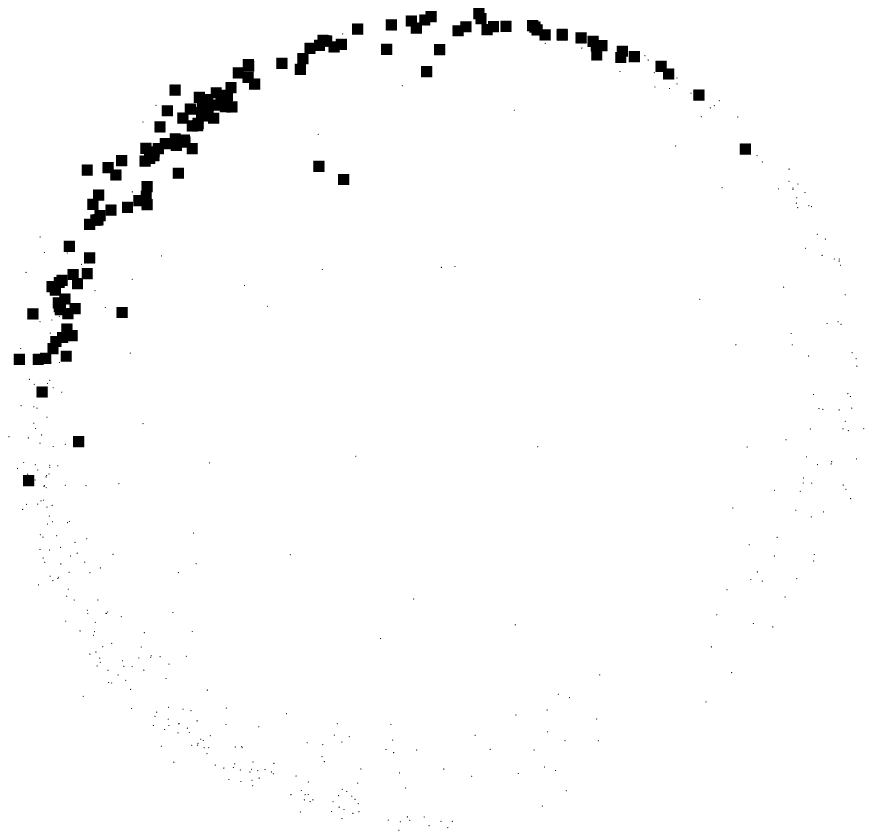}& 
\raisebox{1.5cm}{20 hr} \\

&&\includegraphics[width=3cm]{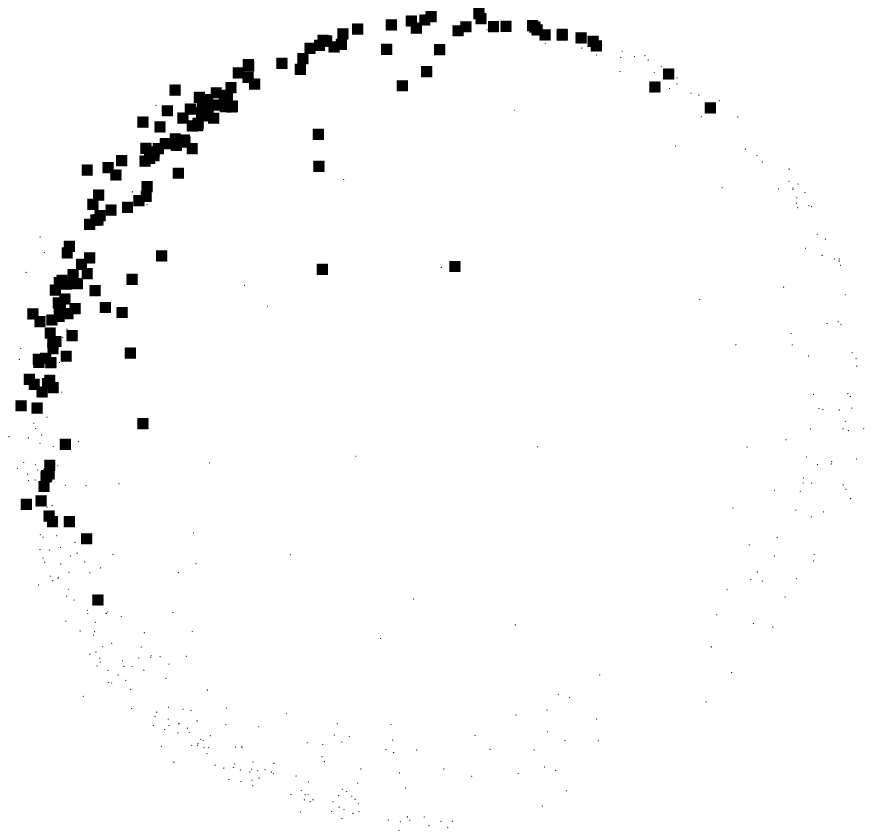}& 
\raisebox{1.5cm}{24 hr} \\
\end{tabular}
\hspace{2cm}
\raisebox{10cm}{(b)}
\hspace{-1cm}
\raisebox{-1cm}{\raisebox{10cm}{time order of cell cycle}}
\hspace{-4cm}
\raisebox{4cm}{\rotatebox{-90}{Angle along ring-like structure [rad]}}
\raisebox{9cm}{\rotatebox{-90}{\includegraphics[width=15cm]{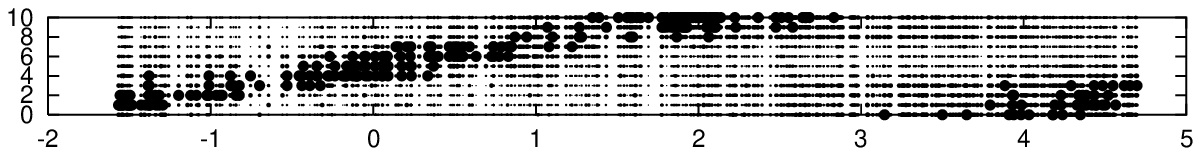}}}

\vspace{-23cm}
Figure 4 (Color) Taguchi \& Oono

\pagebreak

\hfill Figure 5  Taguchi \& Oono

\vspace{-1cm}

\rotatebox{90}{Angle perpendicular to ring-like structure [rad] $\rightarrow$}
\begin{tabular}{rl}
\includegraphics[width=5cm]{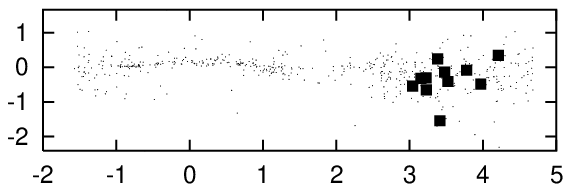} &\raisebox{1cm}{15 min.} \\

\includegraphics[width=5cm]{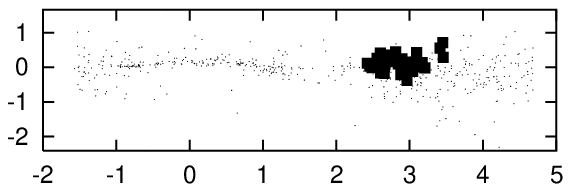}& \raisebox{1cm}{30 min.} \\

\includegraphics[width=5cm]{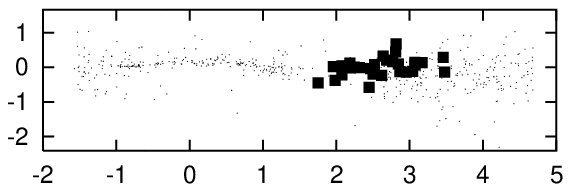}&\raisebox{1cm}{1 hr.} \\

\includegraphics[width=5cm]{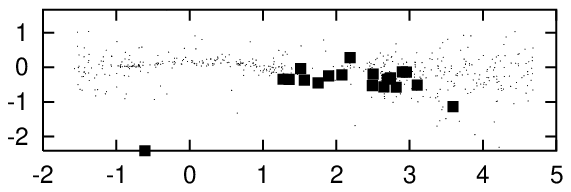}&\raisebox{1cm}{2 hr.}\\

\includegraphics[width=5cm]{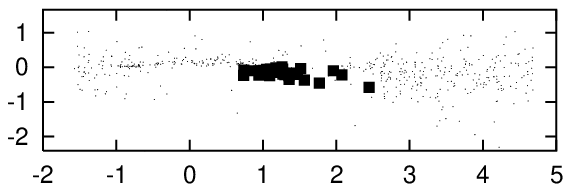}&\raisebox{1cm}{4 hr.}\\

\includegraphics[width=5cm]{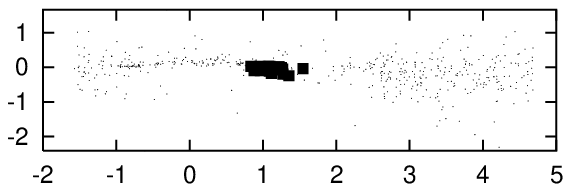}&\raisebox{1cm}{6 hr.}\\

\includegraphics[width=5cm]{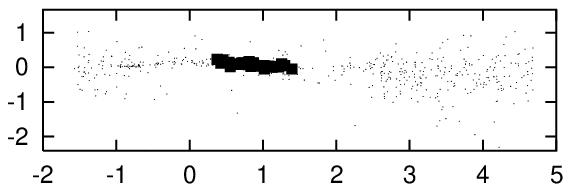}&\raisebox{1cm}{8 hr.}\\

\includegraphics[width=5cm]{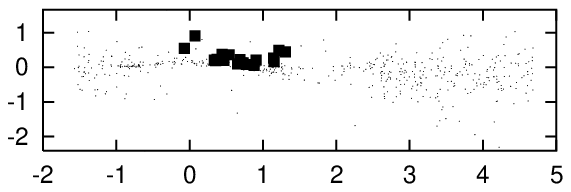}&\raisebox{1cm}{12 hr.}\\

\includegraphics[width=5cm]{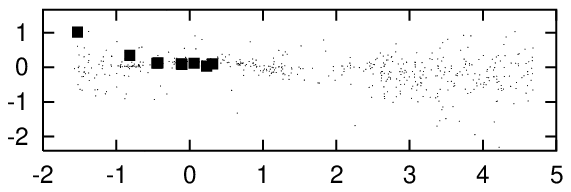}&\raisebox{1cm}{16 hr.}\\

\includegraphics[width=5cm]{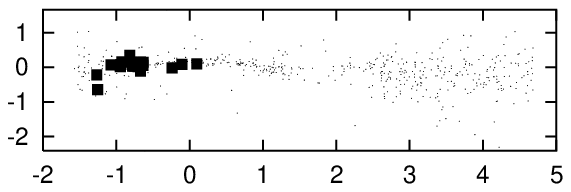}&\raisebox{1cm}{20 hr.}\\

\includegraphics[width=5cm]{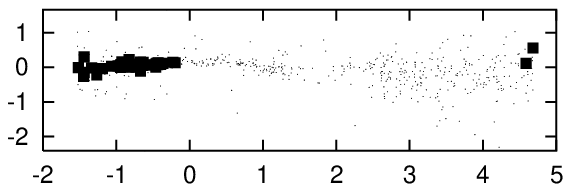}&\raisebox{1cm}{24 hr.}\\
Angle along ring-like structure [rad]  $\rightarrow$
\end{tabular}

\pagebreak
Figure 6  Taguchi \& Oono

\bigskip

\begin{tabular}[b]{lrccc}
&&Data set 1 & Data set 2 & Data set 3 \\
\raisebox{5mm}{\rotatebox{90}{Method 1}}&
\raisebox{5mm}{\rotatebox{90}{\hspace{1cm}$y$}}&
\includegraphics[width=3cm]{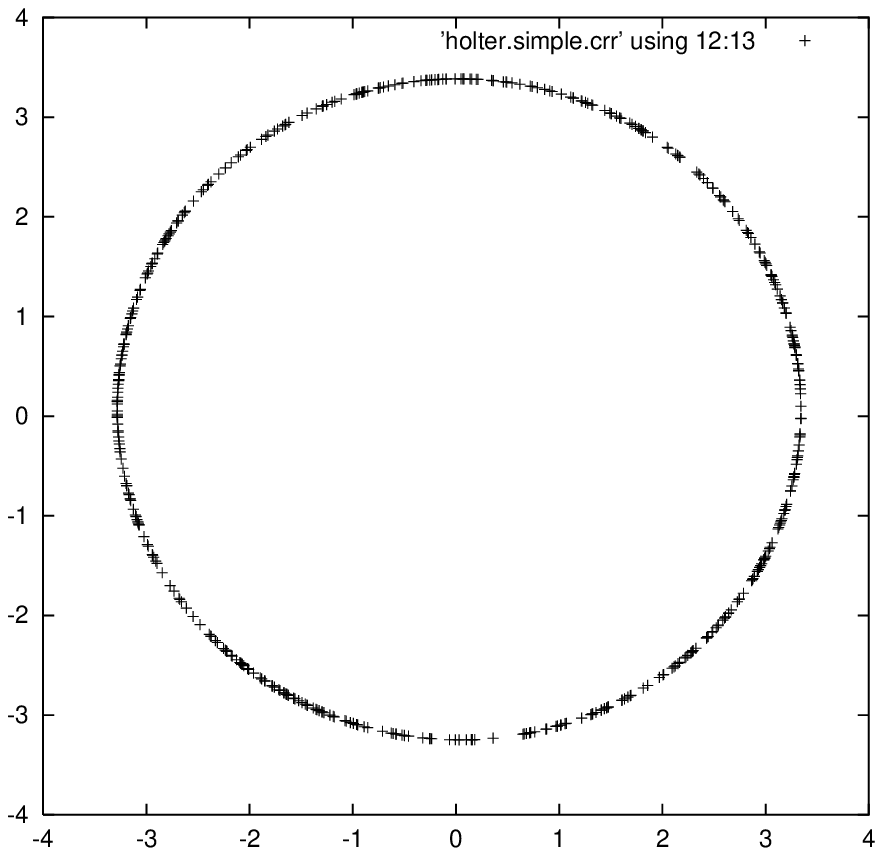}  &
\includegraphics[width=3cm]{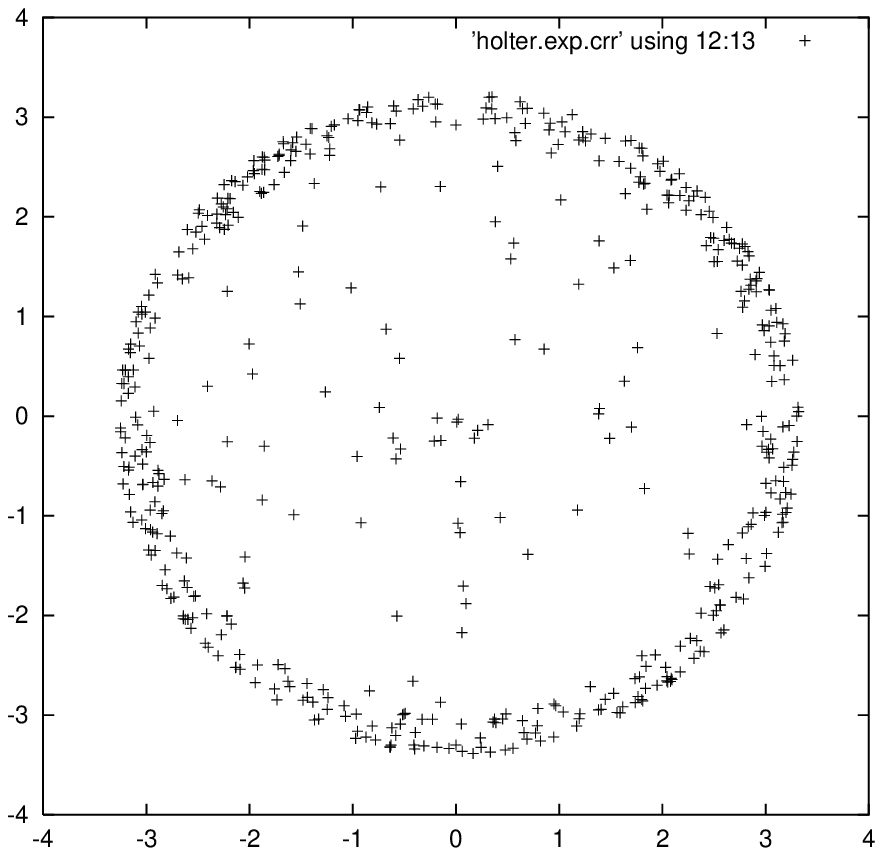} &
\includegraphics[width=3cm]{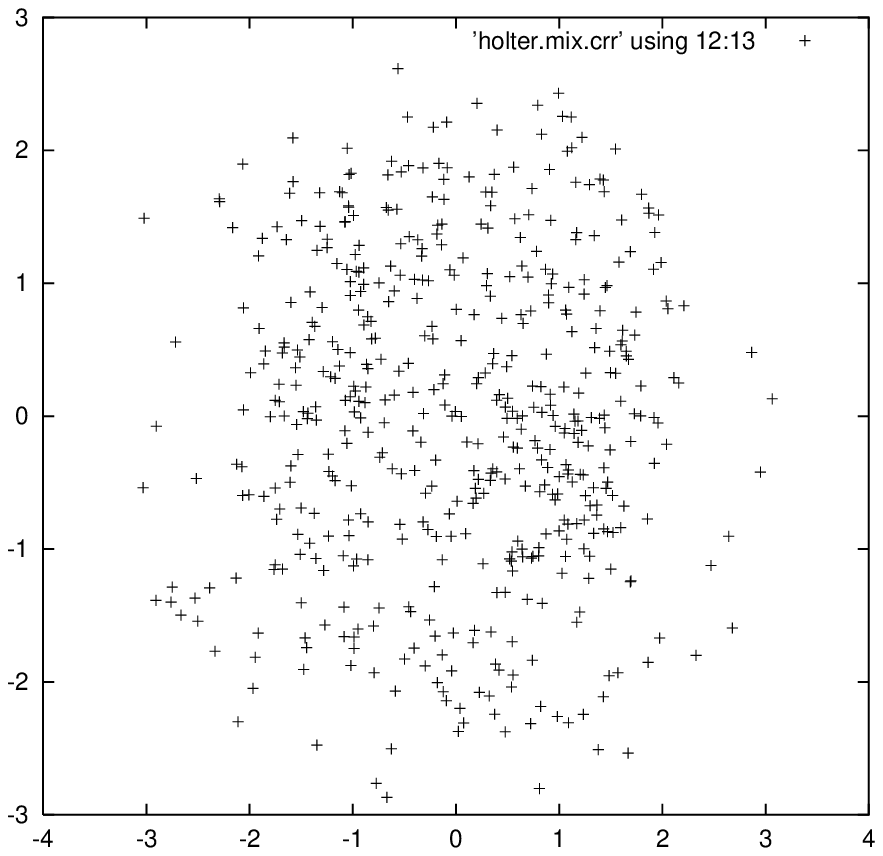}  \\
&& 100 \% & 83 \% & 26\% \\
\raisebox{5mm}{\rotatebox{90}{Method 2}}&
\raisebox{5mm}{\rotatebox{90}{\hspace{1cm}$y$}}&
\includegraphics[width=3cm]{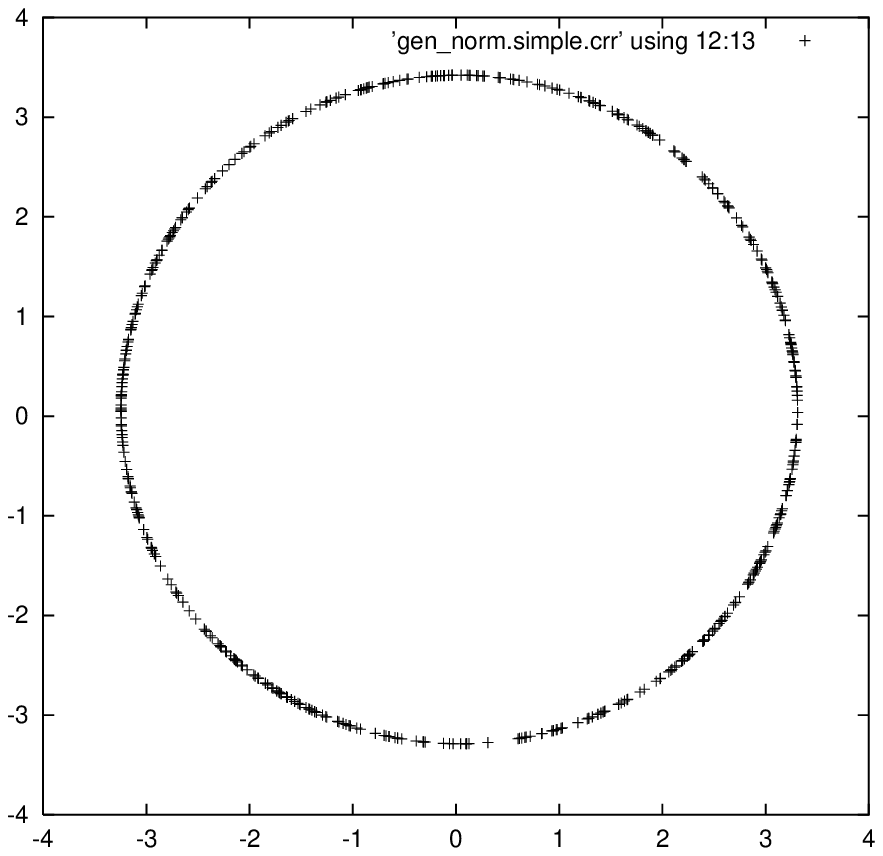}  &
\includegraphics[width=3cm]{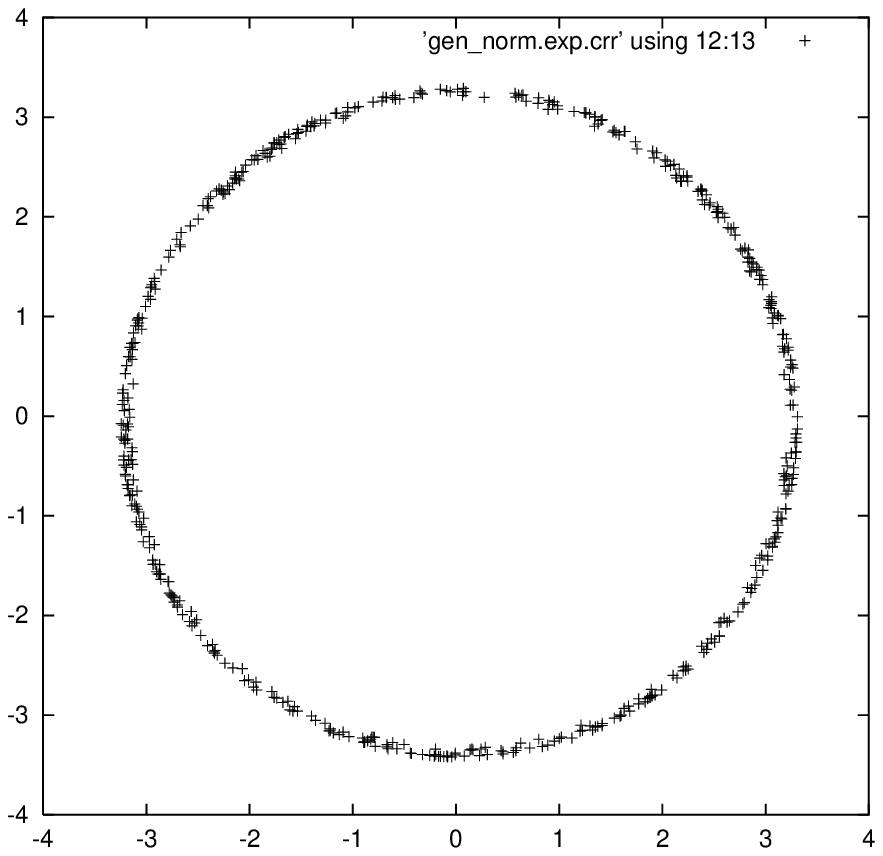} &
\includegraphics[width=3cm]{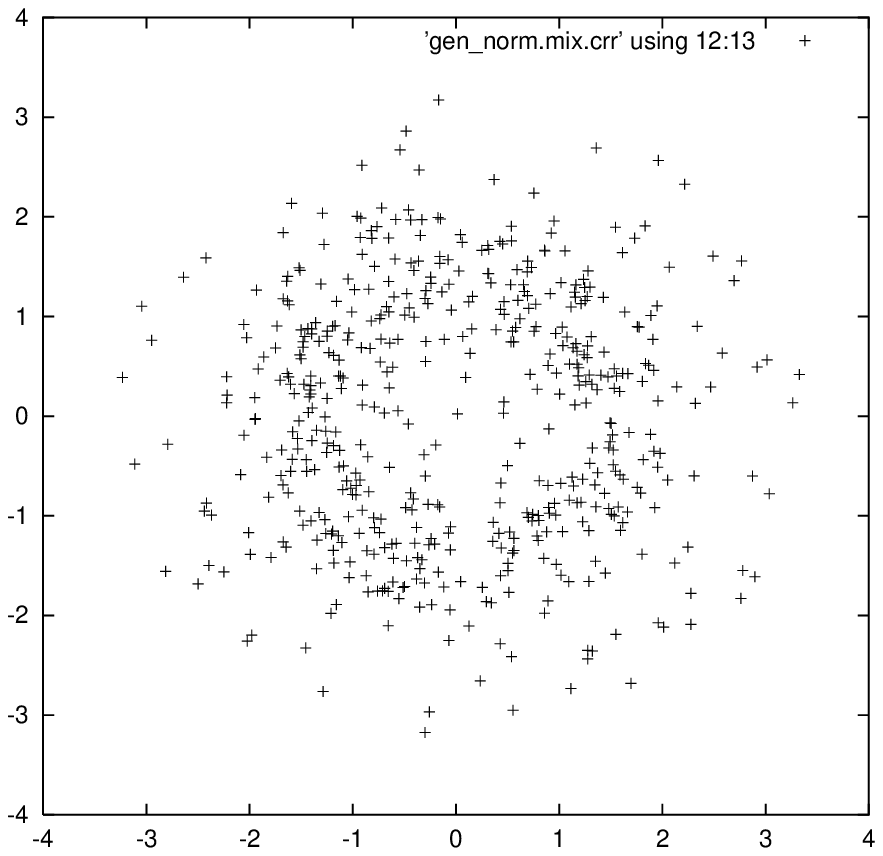} \\
& & 100 \% & 98 \% & 30 \% \\
\raisebox{5mm}{\rotatebox{90}{Method 3}}&
\raisebox{5mm}{\rotatebox{90}{\hspace{1cm}$y$}}&
\includegraphics[width=3cm]{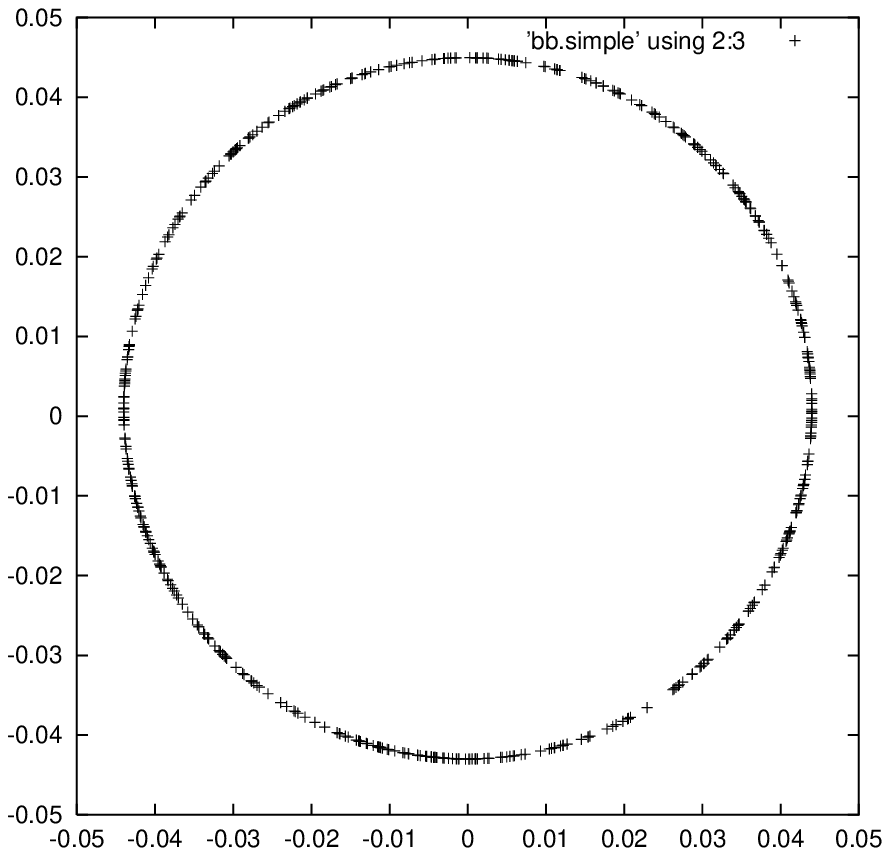}  &
\includegraphics[width=3cm]{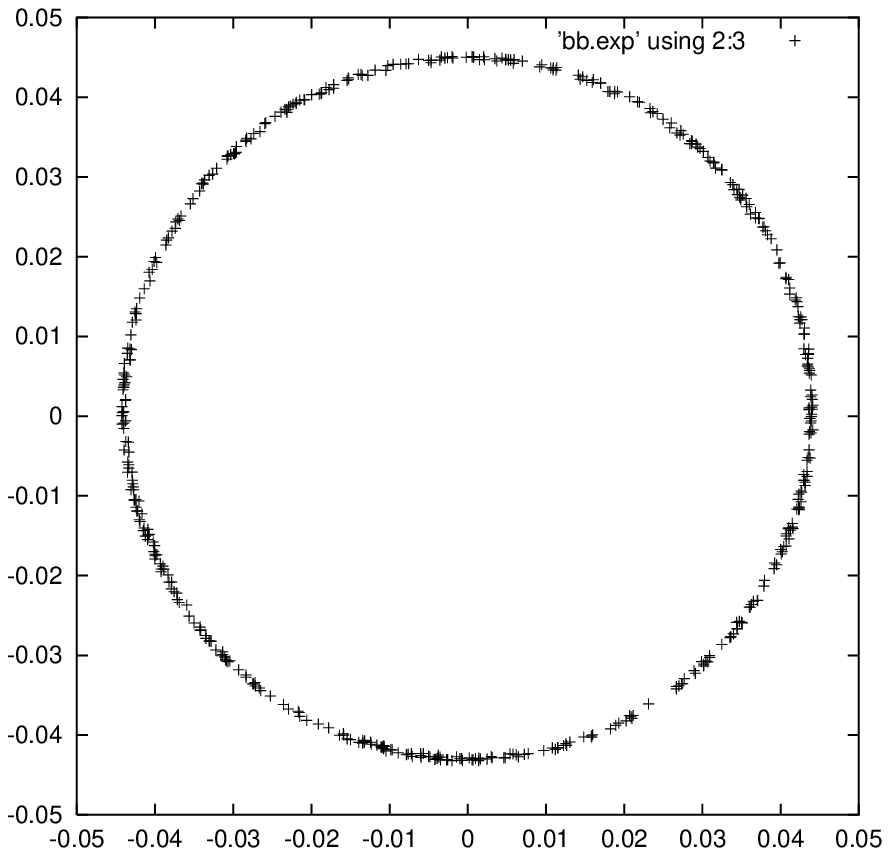} &
\includegraphics[width=3cm]{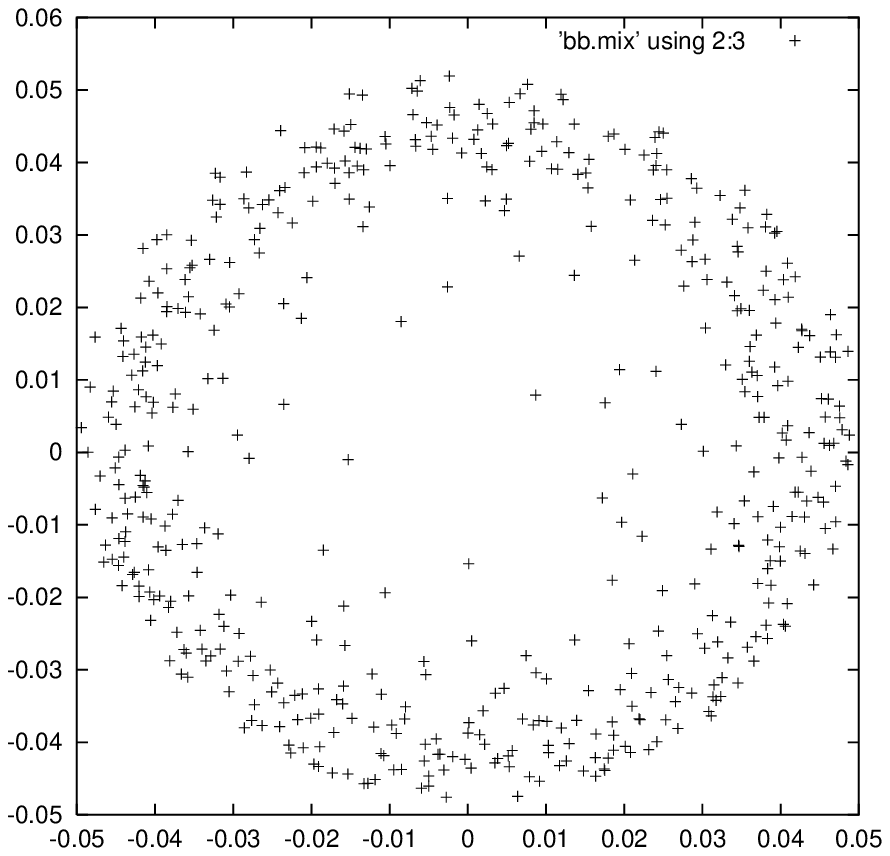} \\
&&  $x$     &   $x$  & $x$  \\
&& 100 \% & 99 \% & 70 \% \\
\end{tabular}

\pagebreak

Figure 7 Taguchi \& Oono

\begin{tabular}{cc}
\rotatebox{90}{\hspace{7cm}\Large{$y$}} &
\includegraphics[scale=1.5]{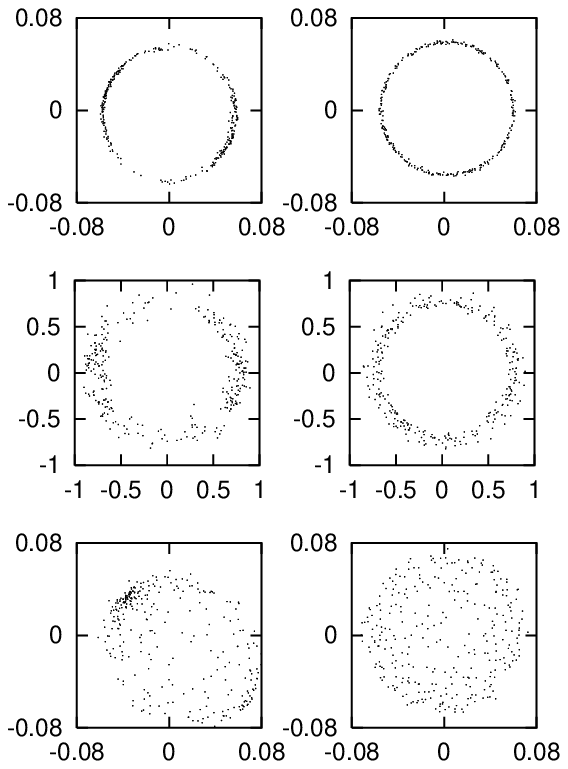}\\
      & \hspace{-8cm} \Large{$x$} \\
\end{tabular}

\pagebreak
\noindent
Figure legends for online only color figures:

\noindent
Figure 1 (Color; online only):
PCA results using correlation coefficient matrix.
The first two principal components are used as the horizontal and
vertical axis, respectively (the cumulative proportion is 70 \%).
Colors indicate relative intensity of experimental values
(red $> 3.2$, yellow $> 2.4$, green $> 1.6$, pale blue $> 0.8$,
gray $< 0.8$).
       From the top the time is, respectively, 15 min., 30 min., 1 hr, 2 hr,
4 hr, 6 hr, 8 hr, 12 hr, 16 hr, 20 hr, 24 hr.

\noindent
Figure 4 (Color; online only):
(a) Temporal patterns of gene expression levels visualized with the aid of
nMDS.
Colors indicate relative intensity of experimental values
normalized so that $\sum_t s_{gt}=0$
and $\sum_t s_{gt}^2 =1$ where $s_{gt}$ is experimental variable
of $g$th genes at time $t$.
(red $> 1.6$, yellow $> 1.2$, green $> 0.8$, pale blue $> 0.4$,
gray $< 0.4$).
Time sequences are the same as explained at Fig.\ 1.
\noindent
(b) Gene expression data as a function of the angle measured from the
vertical axis in (a).
The
horizontal axis corresponds to $t$. The color convention is the
same as in (a).  The figures are arranged in two columns, but this
is solely for the layout purpose; there is no distinction between
two columns.

\noindent
Figure 5 (Color: online only):
3D unfolding of the temporal pattern of gene expression level with the aid of
nMDS (3D).
The color convention is the same as explained in Figure 3.
The horizontal (resp., vertical) axis represents $\phi$ (resp., $\theta$).
See the text for detail.   The figures are arranged in two columns, but this
is solely for the layout purpose; there is no distinction between
two columns.

\pagebreak
\hfill Figure 1 (Color) Taguchi \& Oono

\vspace{-2cm}
\begin{tabular} {rccl}

\raisebox{1.5cm}{15 min.}& \includegraphics[width=3cm]{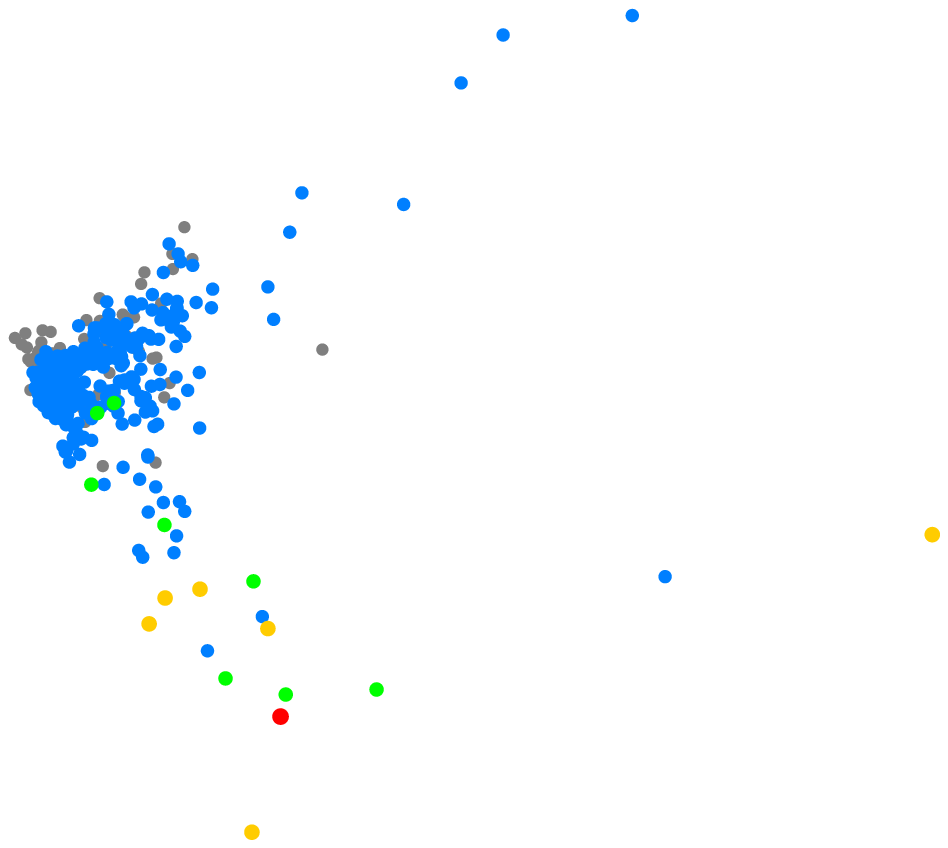} &&\\

\raisebox{1.5cm}{30 min.}& \includegraphics[width=3cm]{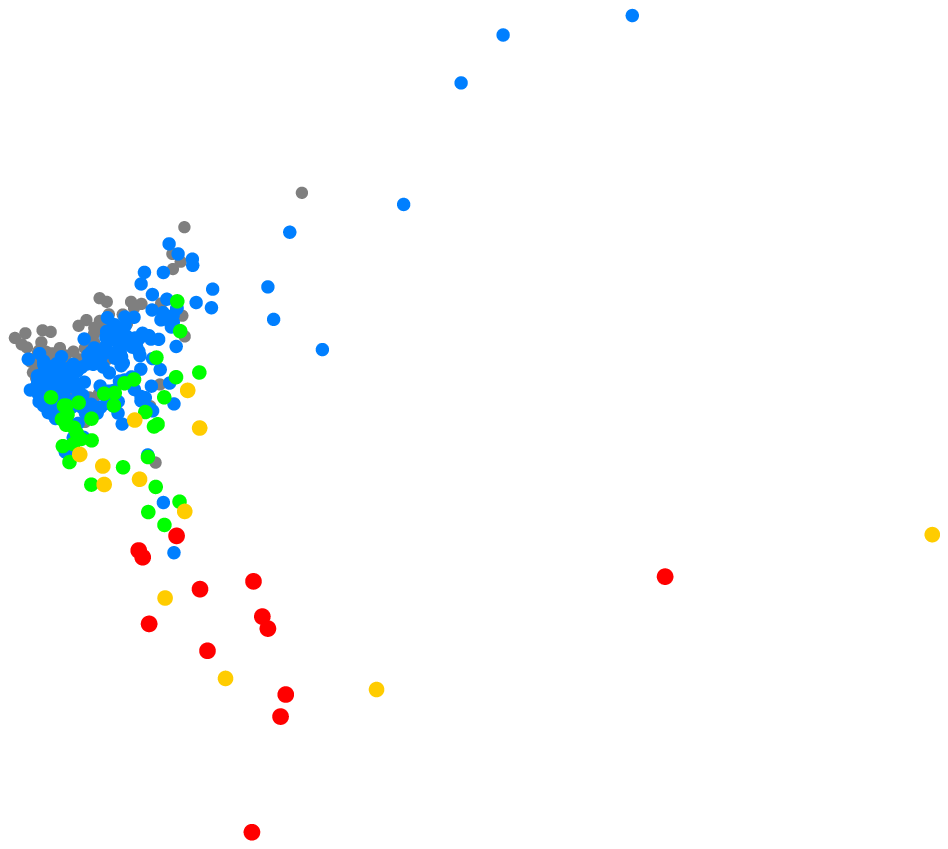} &
\includegraphics[width=3cm]{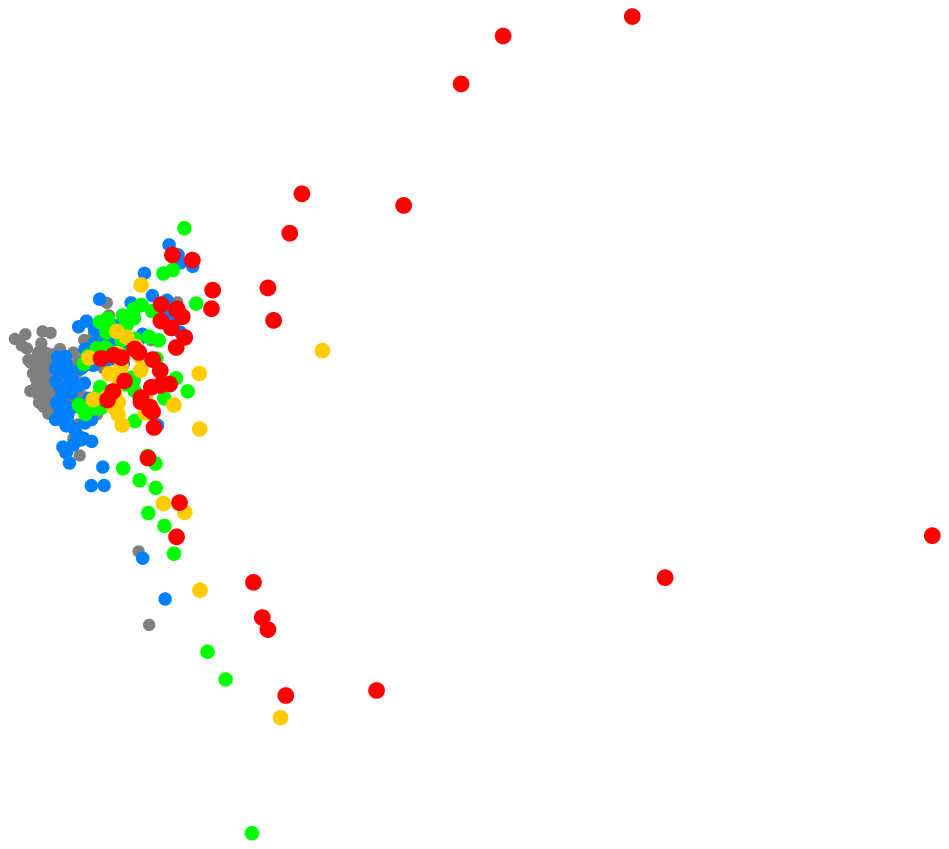} & \raisebox{1.5cm}{6 hr}\\

\raisebox{1.5cm}{1 hr}& \includegraphics[width=3cm]{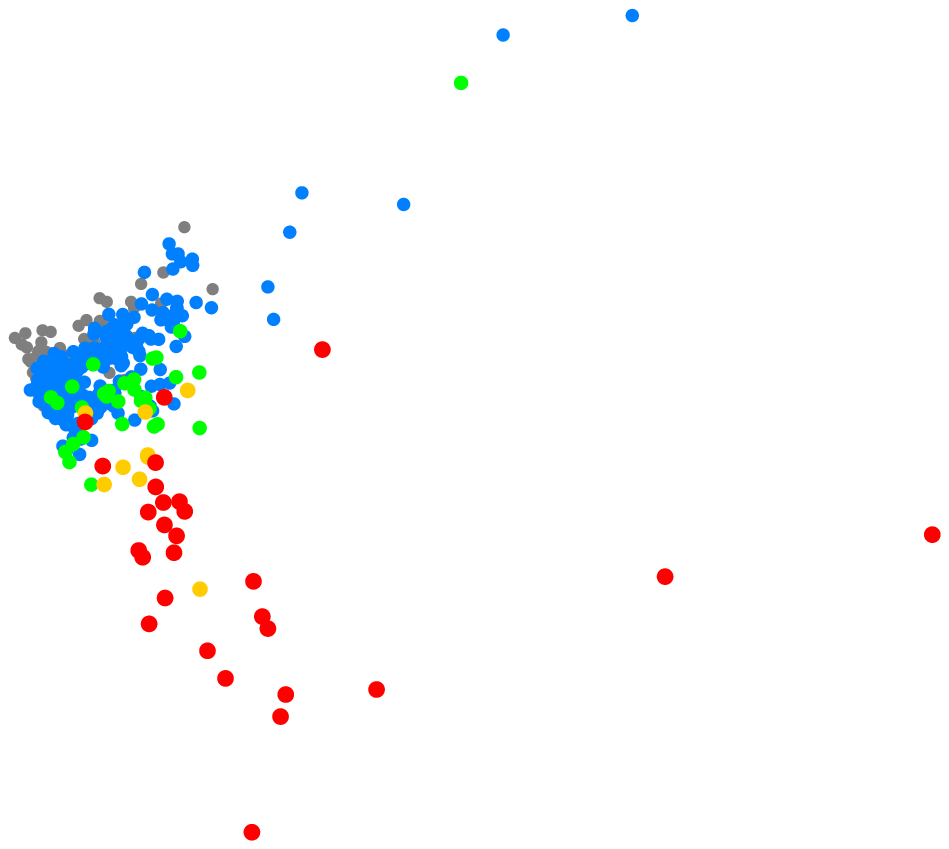} &
\includegraphics[width=3cm]{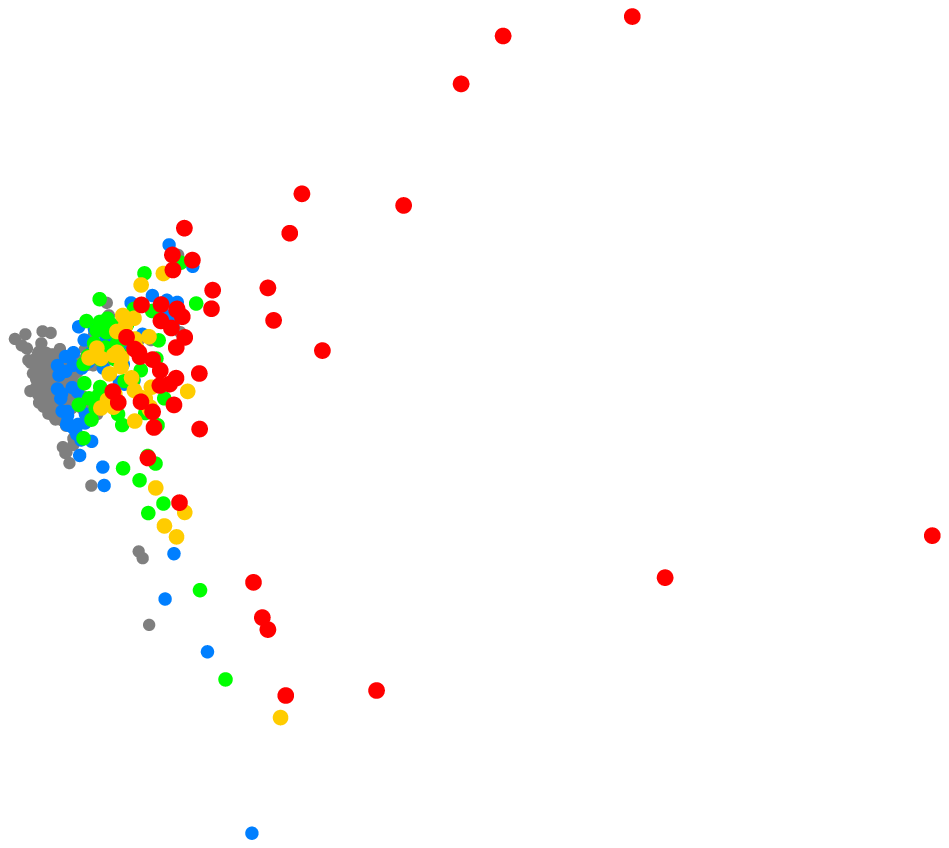}& \raisebox{1.5cm}{8 hr} \\

\raisebox{1.5cm}{2 hr}& \includegraphics[width=3cm]{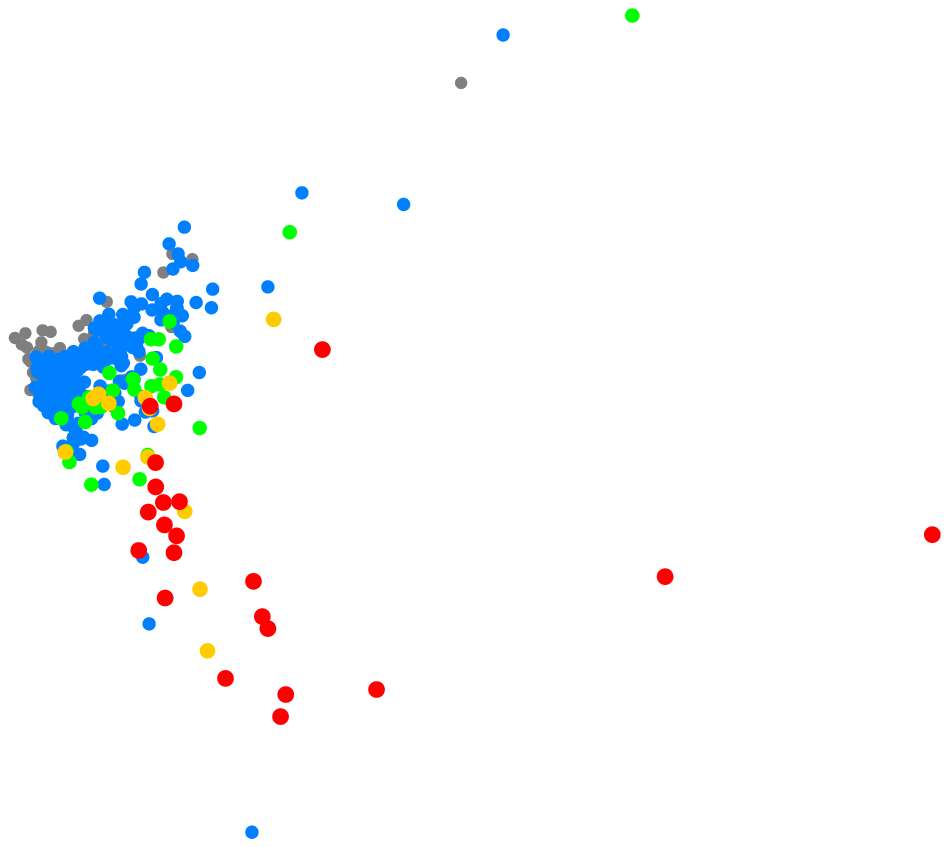} &
\includegraphics[width=3cm]{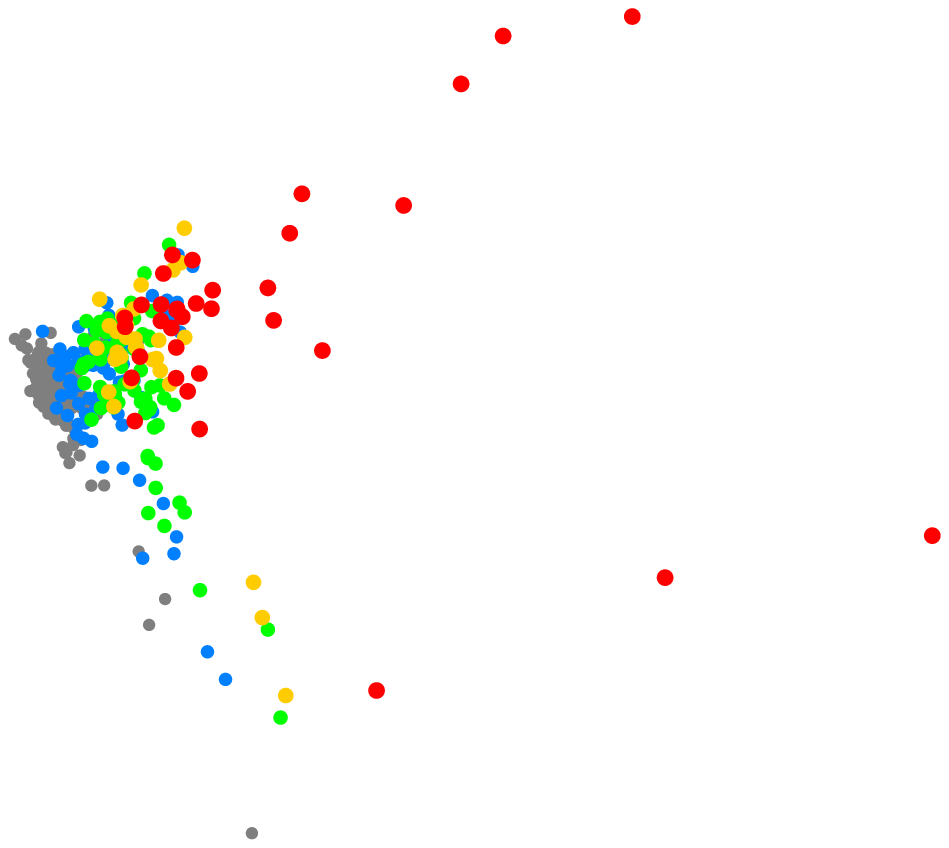}& \raisebox{1.5cm}{12 hr}\\

\raisebox{1.5cm}{4 hr}& \includegraphics[width=3cm]{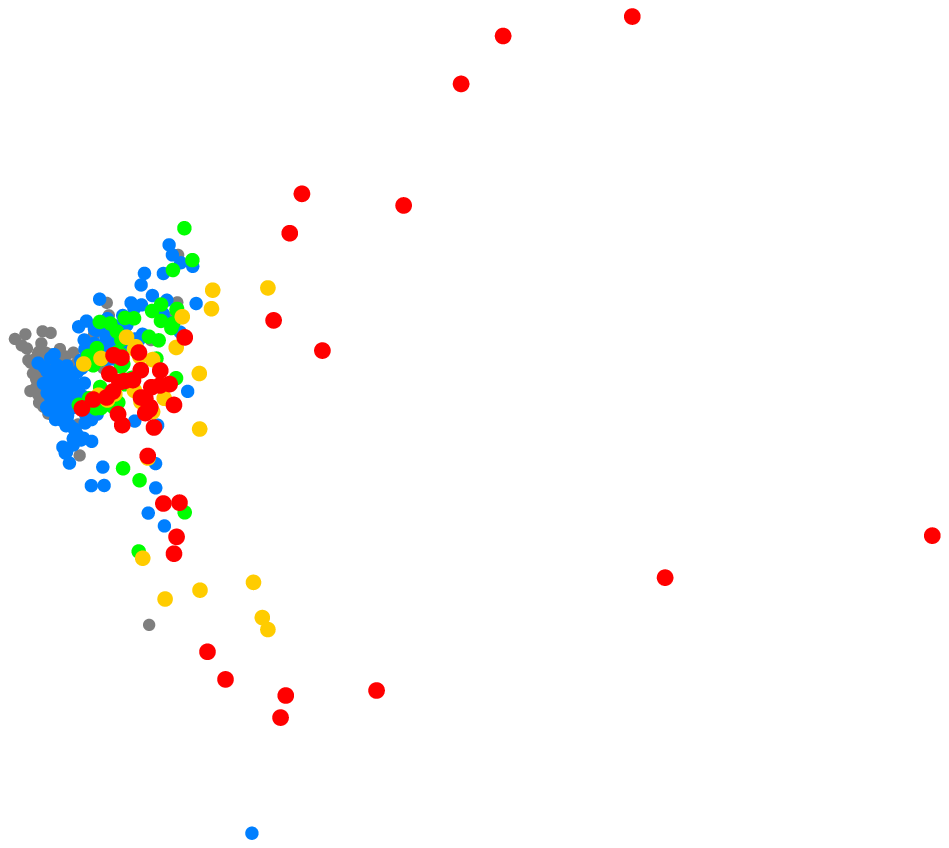} &
\includegraphics[width=3cm]{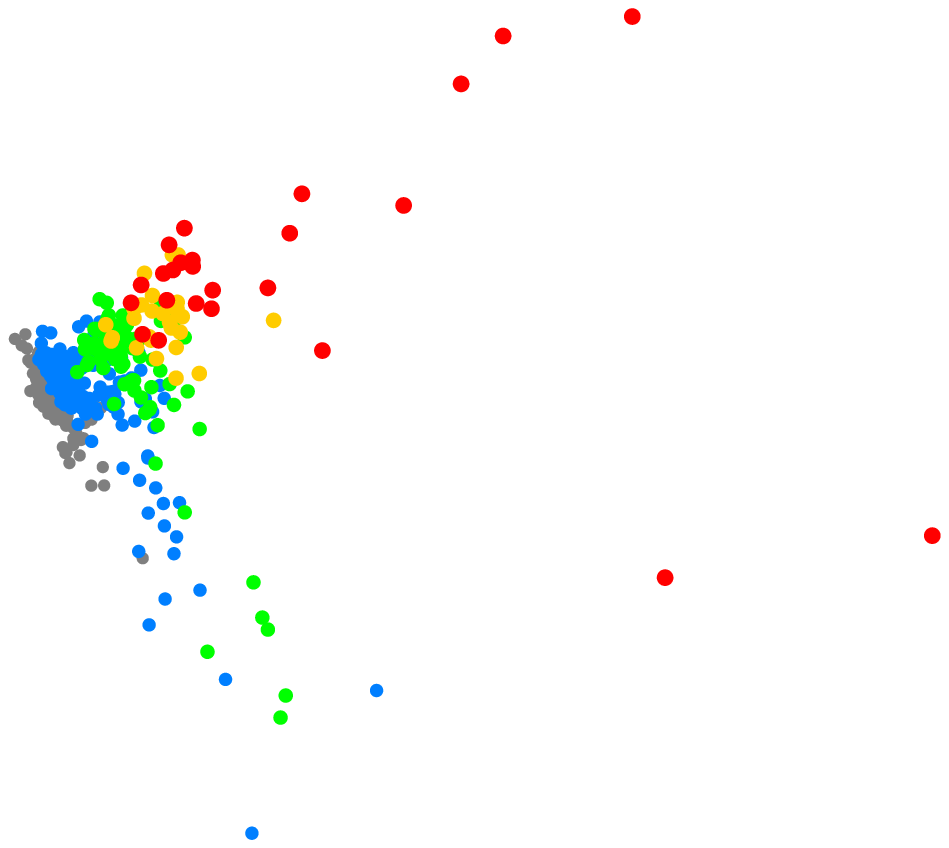} & \raisebox{1.5cm}{16 hr} \\





& & \includegraphics[width=3cm]{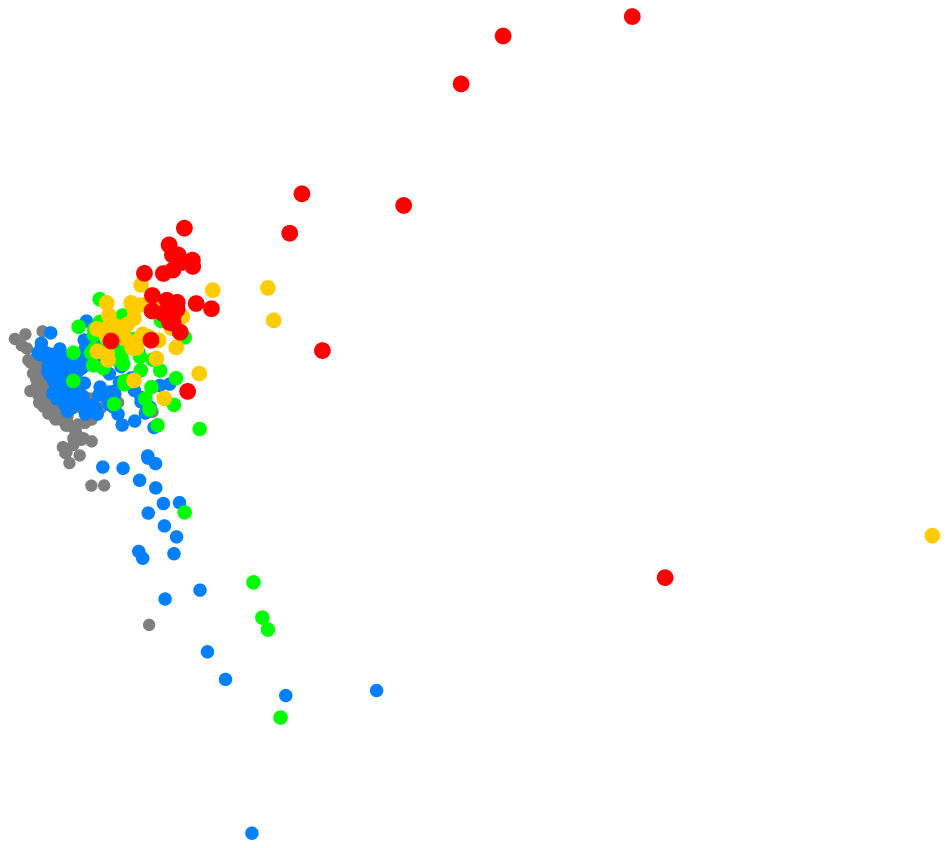}& \raisebox{1.5cm}{20 hr} \\

&&\includegraphics[width=3cm]{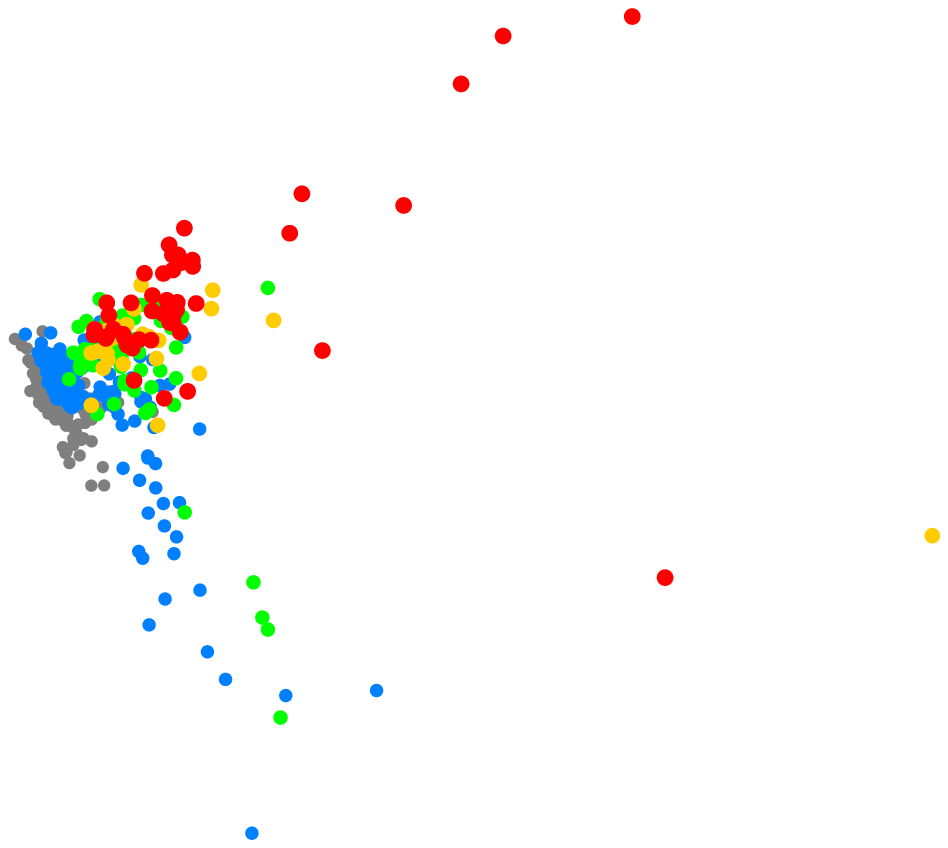}& \raisebox{1.5cm}{24 hr} \\
\end{tabular}

\pagebreak


\begin{tabular} {rccl}

\raisebox{3cm}{(a)}
\raisebox{1.5cm}{15 min.}& \includegraphics[width=3cm]{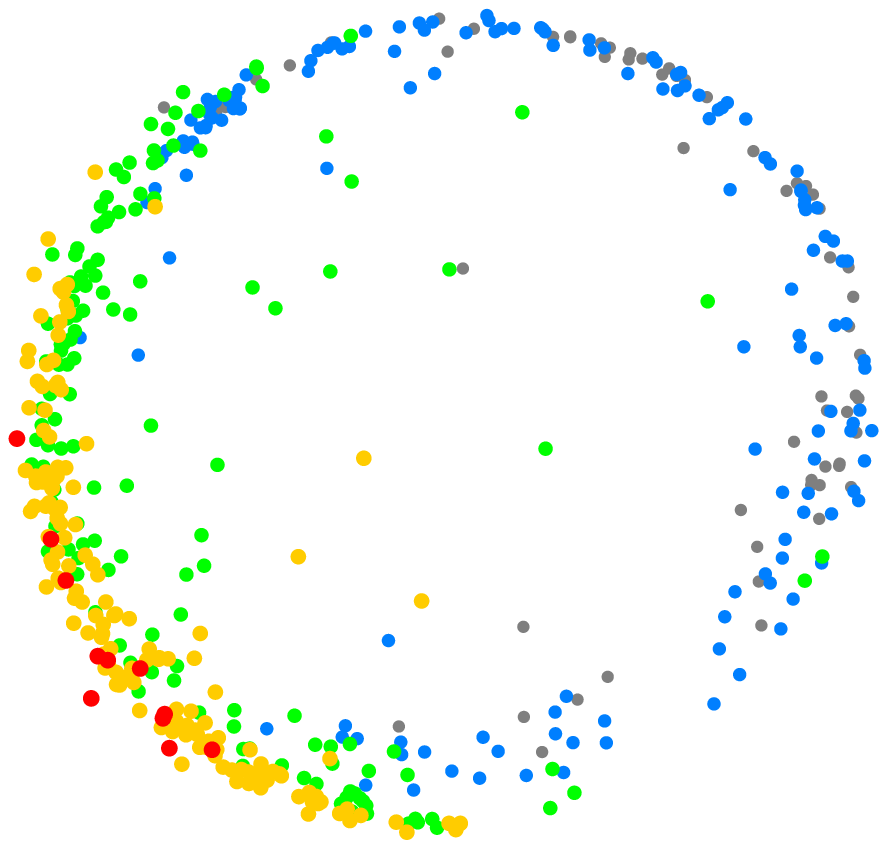} &&\\

\raisebox{1.5cm}{30 min.}& \includegraphics[width=3cm]{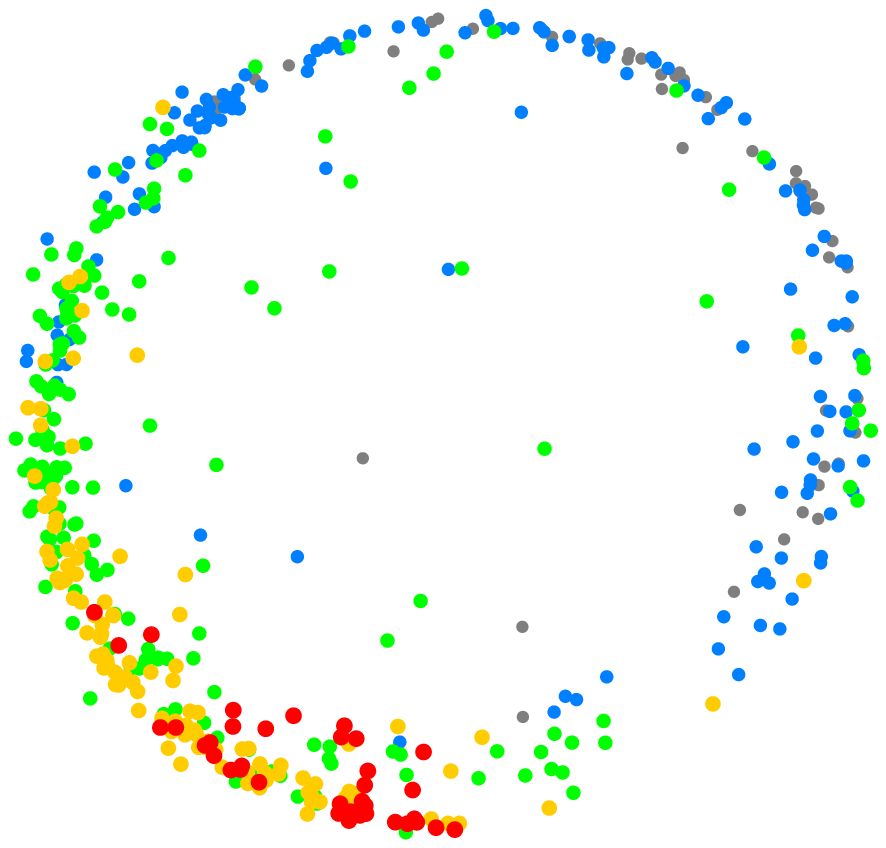} &
\includegraphics[width=3cm]{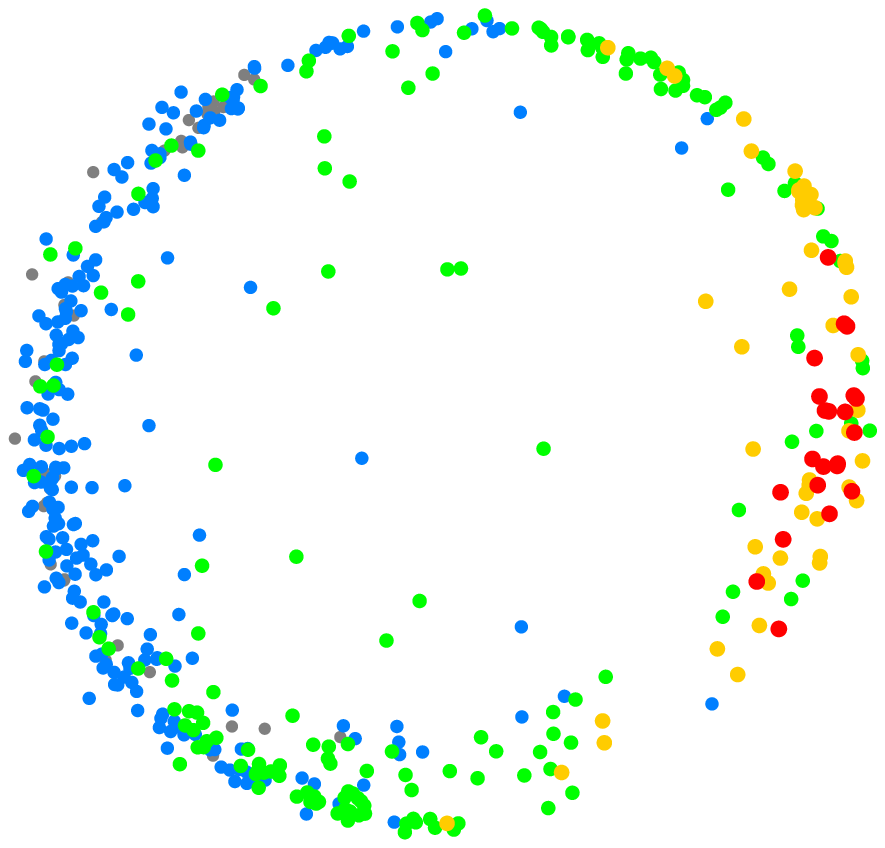} & \raisebox{1.5cm}{6 hr}\\

\raisebox{1.5cm}{1 hr}& \includegraphics[width=3cm]{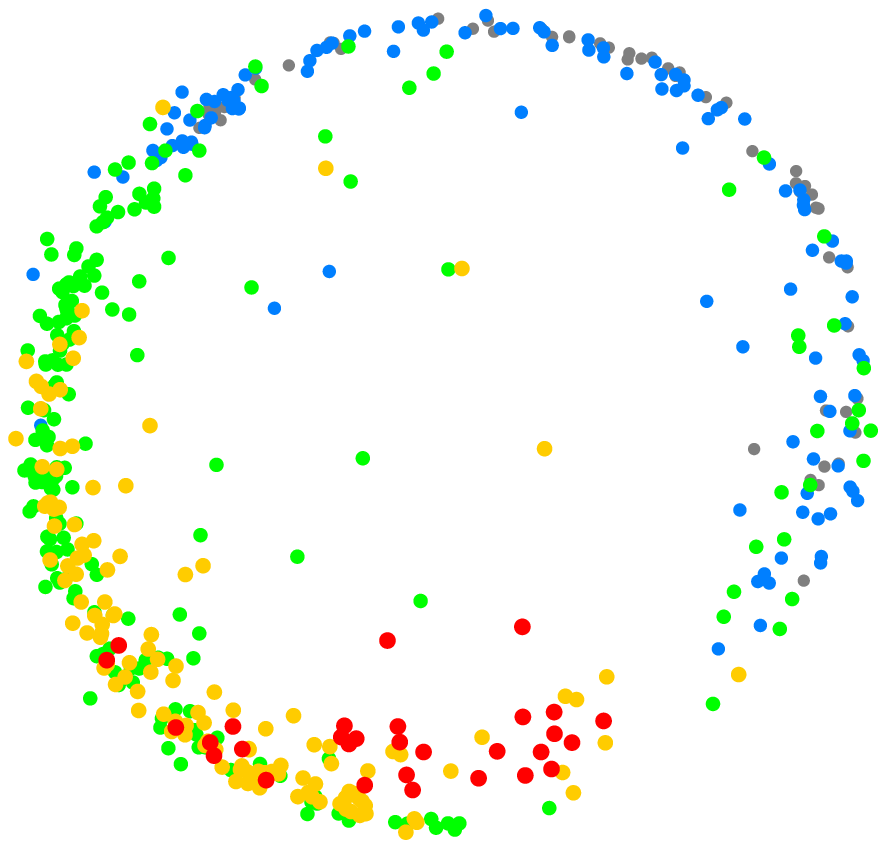} &
\includegraphics[width=3cm]{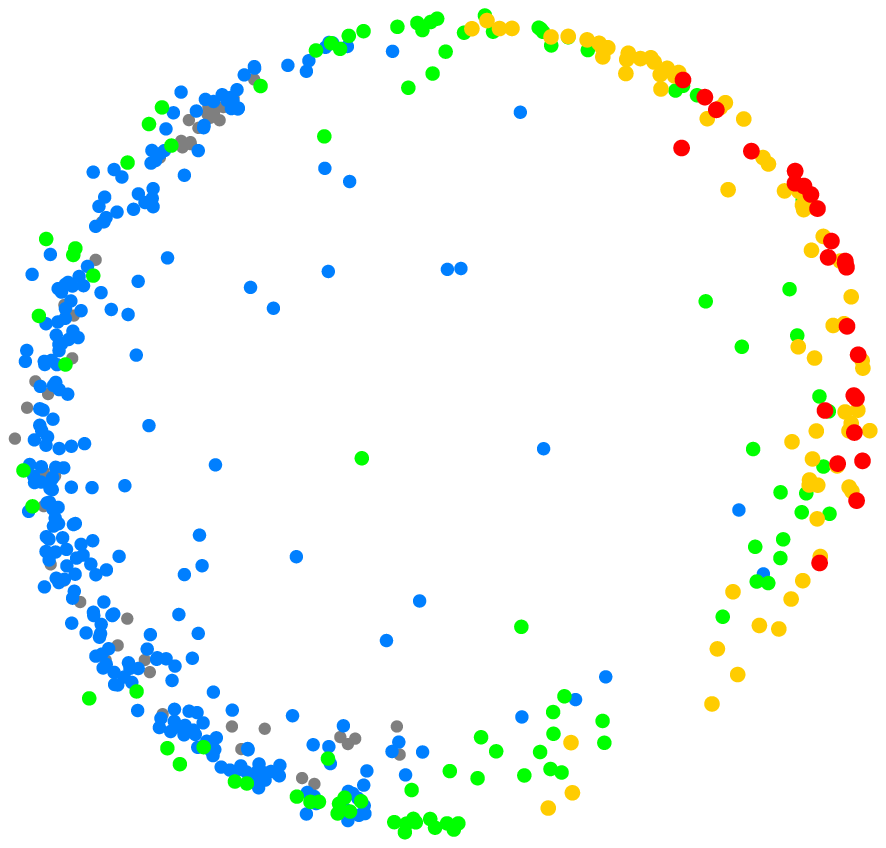}& \raisebox{1.5cm}{8 hr} \\

\raisebox{1.5cm}{2 hr}& \includegraphics[width=3cm]{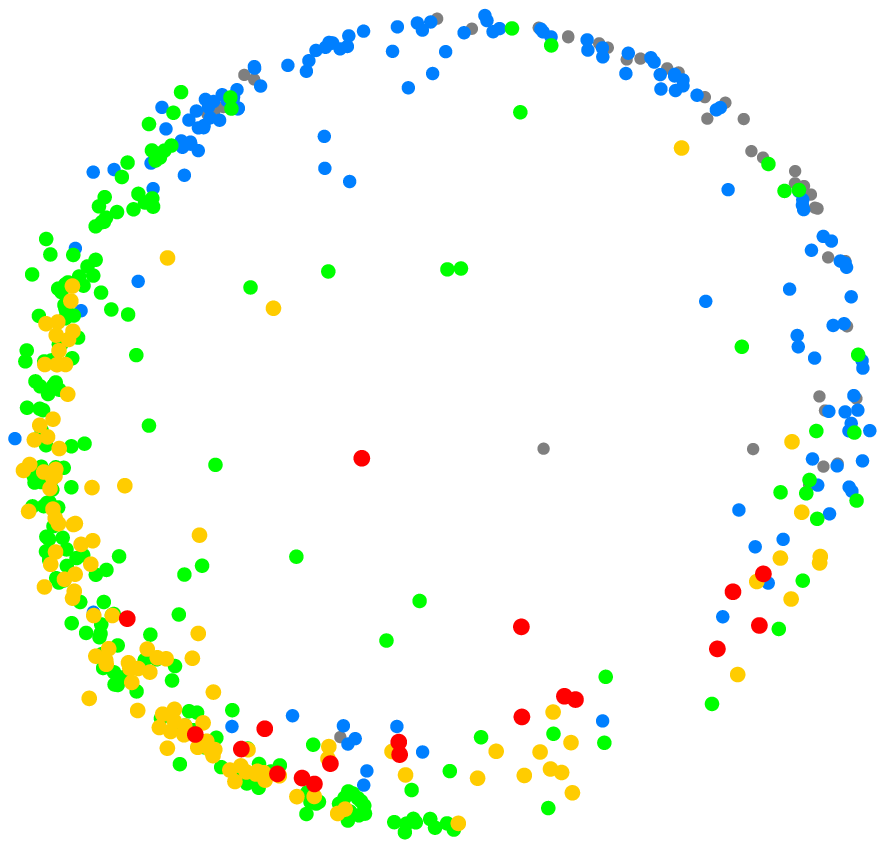} &
\includegraphics[width=3cm]{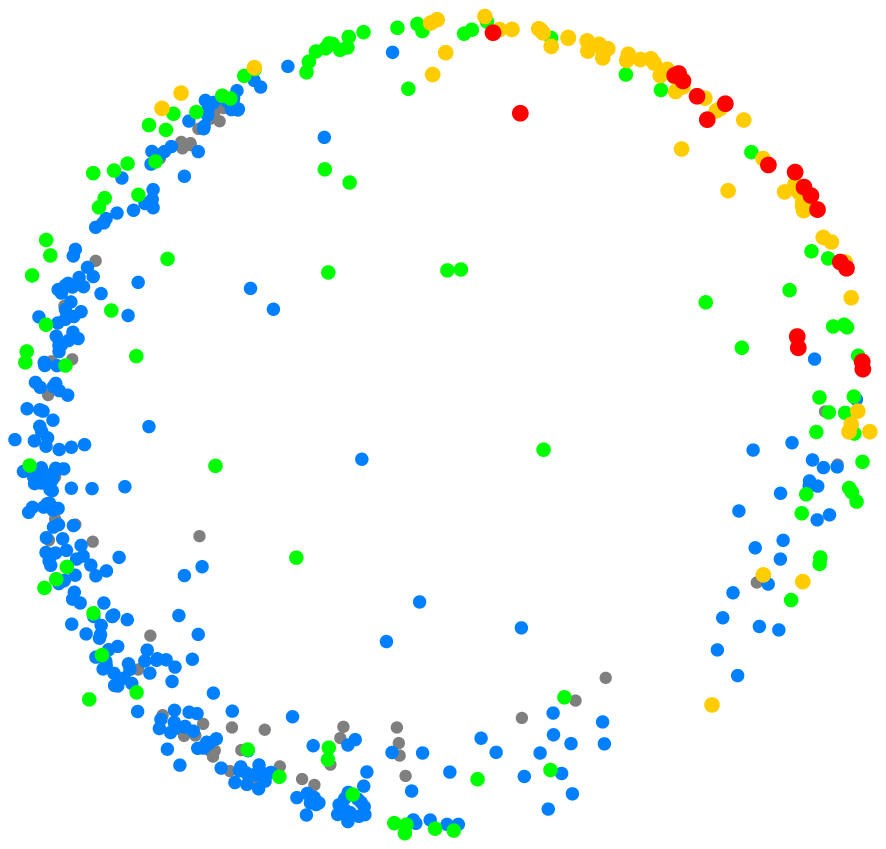}&\raisebox{1.5cm}{12 hr}\\

\raisebox{1.5cm}{4 hr}& \includegraphics[width=3cm]{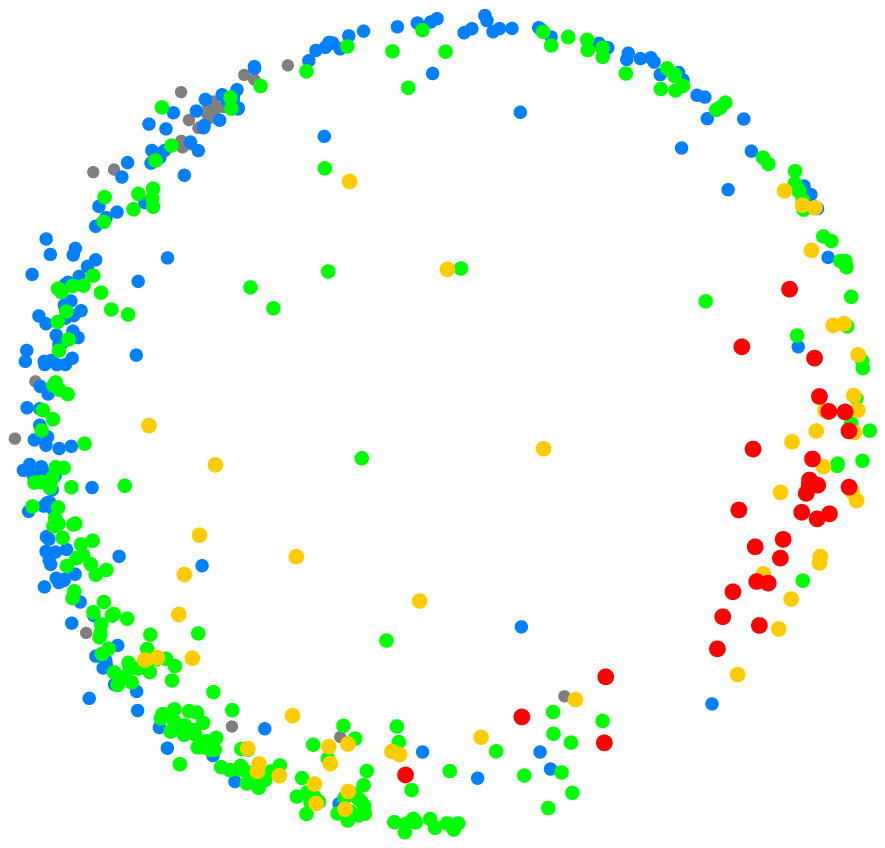} &
\includegraphics[width=3cm]{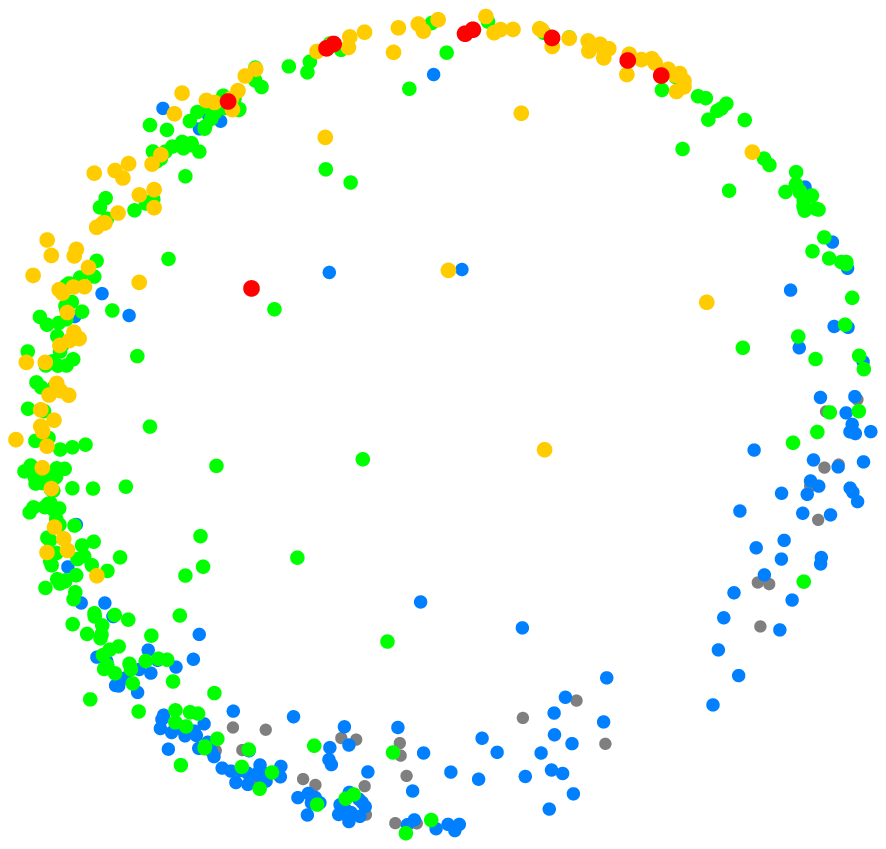} & \raisebox{1.5cm}{16 hr} \\





& & \includegraphics[width=3cm]{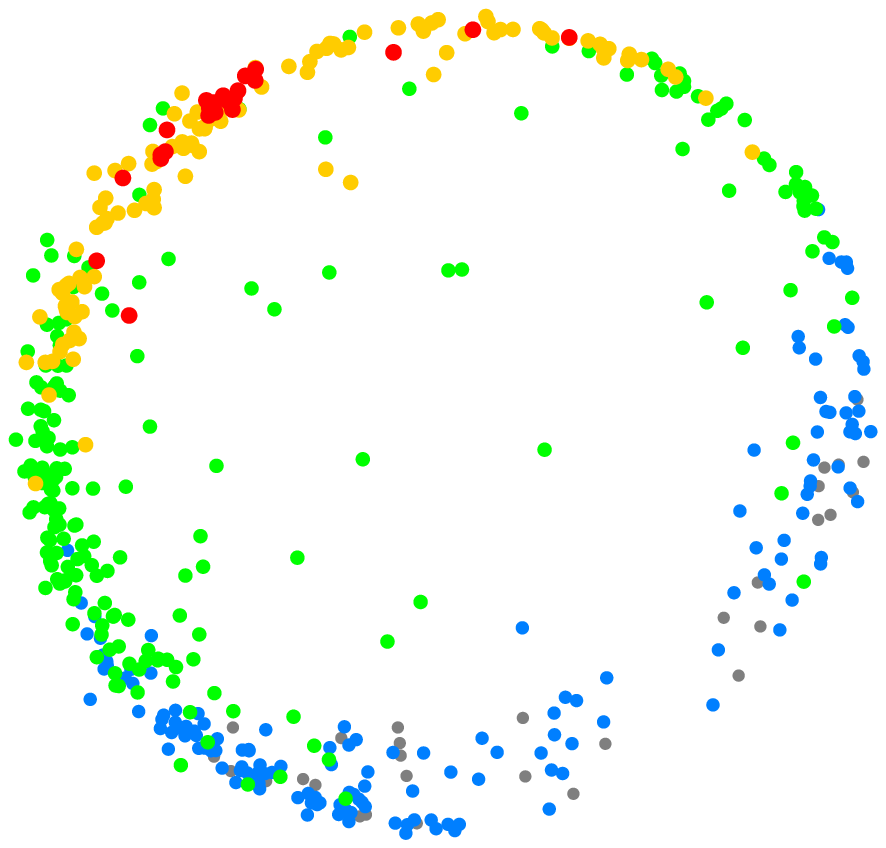}& \raisebox{1.5cm}{20 hr} \\

&&\includegraphics[width=3cm]{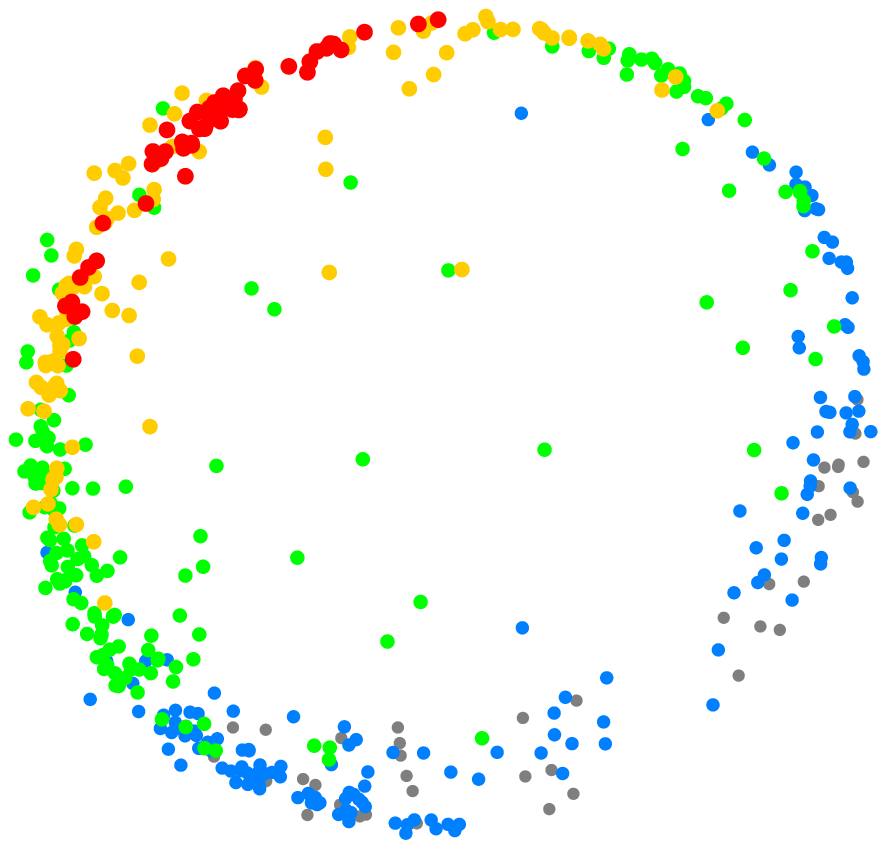}& \raisebox{1.5cm}{24 hr} \\
\end{tabular}
\hspace{2cm}
\raisebox{10cm}{(b)}
\hspace{-1cm}
\raisebox{-1cm}{\raisebox{10cm}{time order of cell cycle}}
\hspace{-4cm}
\raisebox{4cm}{\rotatebox{-90}{Angle along ring-like structure [rad]}}
\raisebox{9cm}{\rotatebox{-90}{\includegraphics[width=15cm]{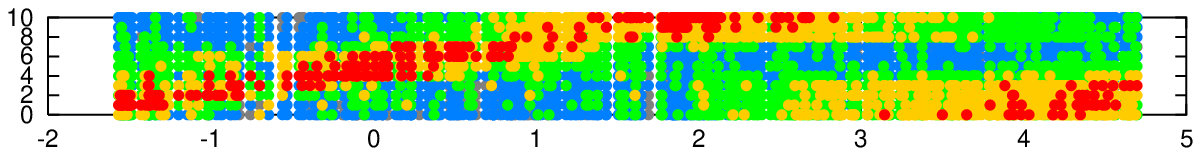}}}

\vspace{-23cm}
Figure 4 (Color) Taguchi \& Oono

\pagebreak

\hfill Figure 5 (Color)  Taguchi \& Oono

\vspace{-1cm}

\rotatebox{90}{Angle perpendicular to ring-like structure [rad] $\rightarrow$}
\begin{tabular}{rl}
\includegraphics[width=5cm]{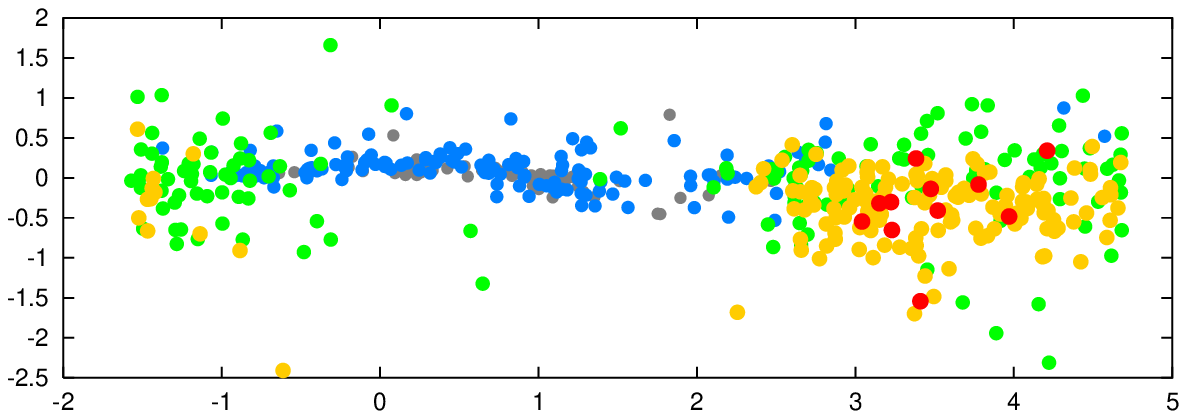} &\raisebox{1cm}{15 min.} \\

\includegraphics[width=5cm]{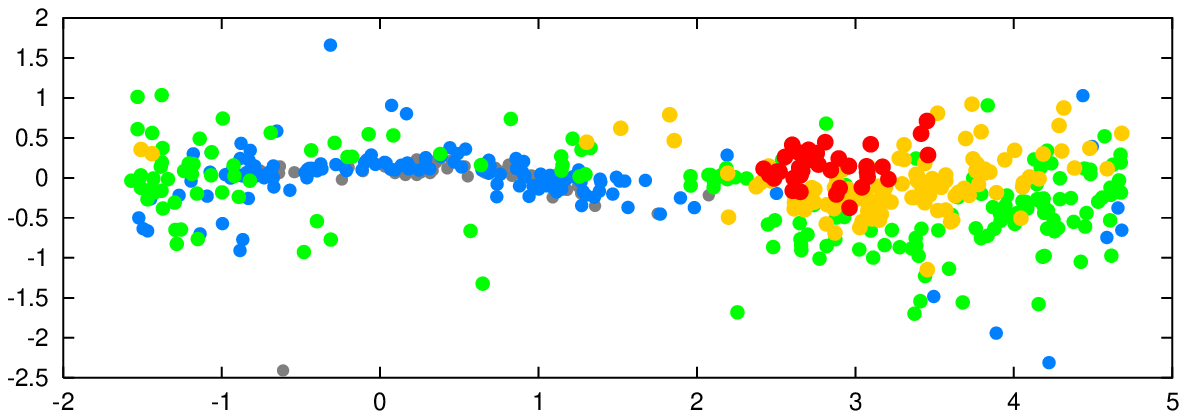}& \raisebox{1cm}{30 min.} \\

\includegraphics[width=5cm]{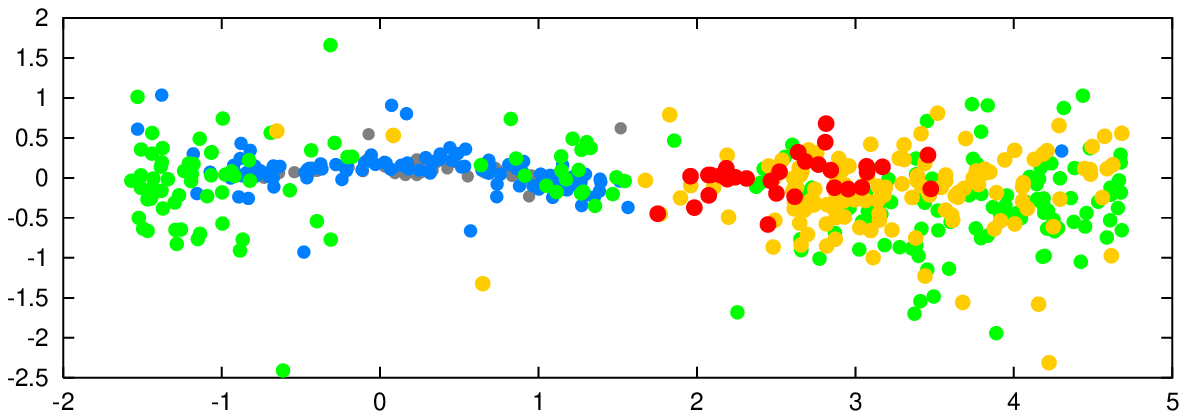}&\raisebox{1cm}{1 hr.} \\

\includegraphics[width=5cm]{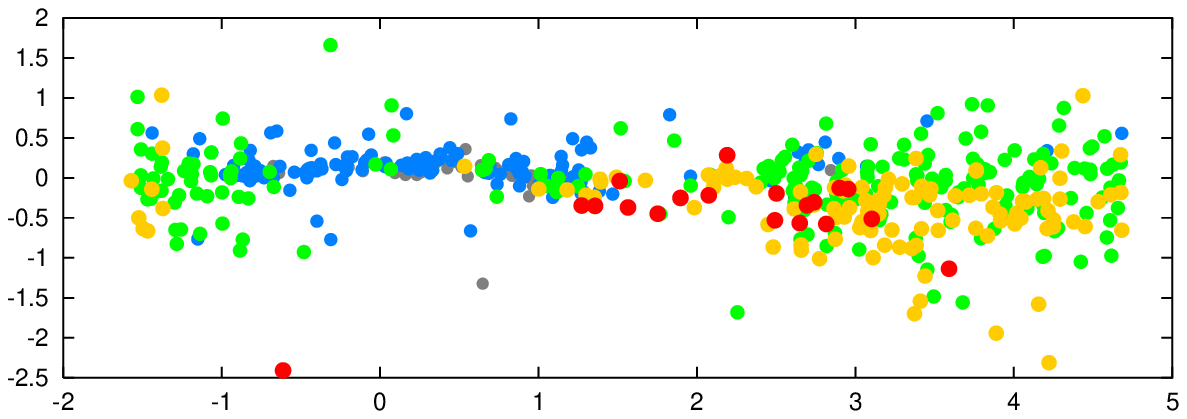}&\raisebox{1cm}{2 hr.}\\

\includegraphics[width=5cm]{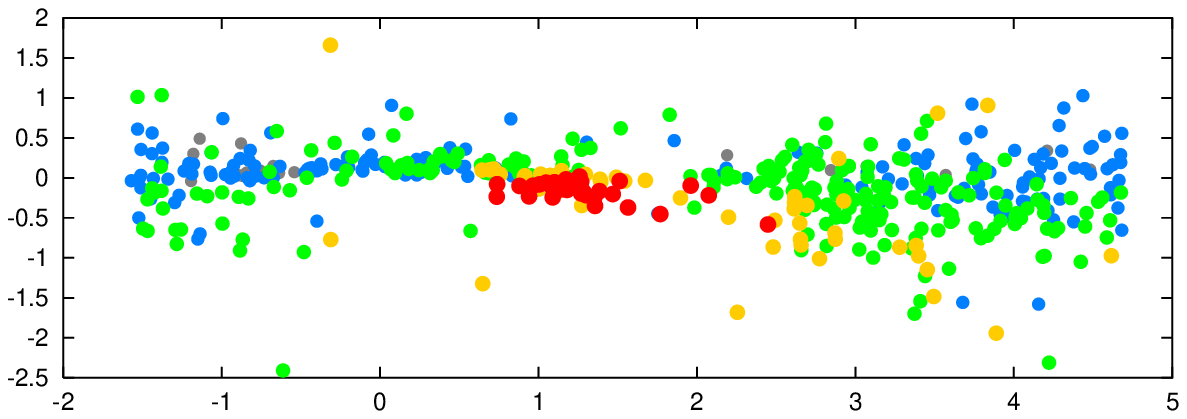}&\raisebox{1cm}{4 hr.}\\

\includegraphics[width=5cm]{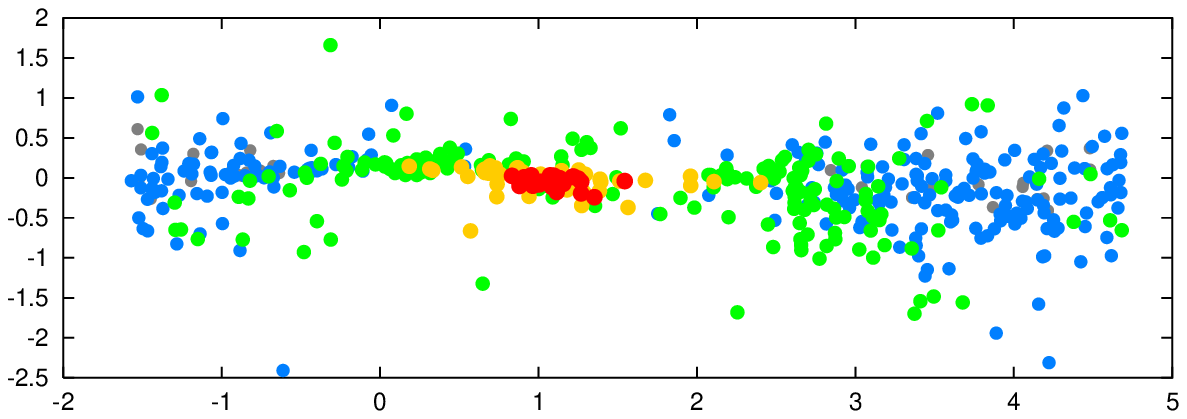}&\raisebox{1cm}{6 hr.}\\

\includegraphics[width=5cm]{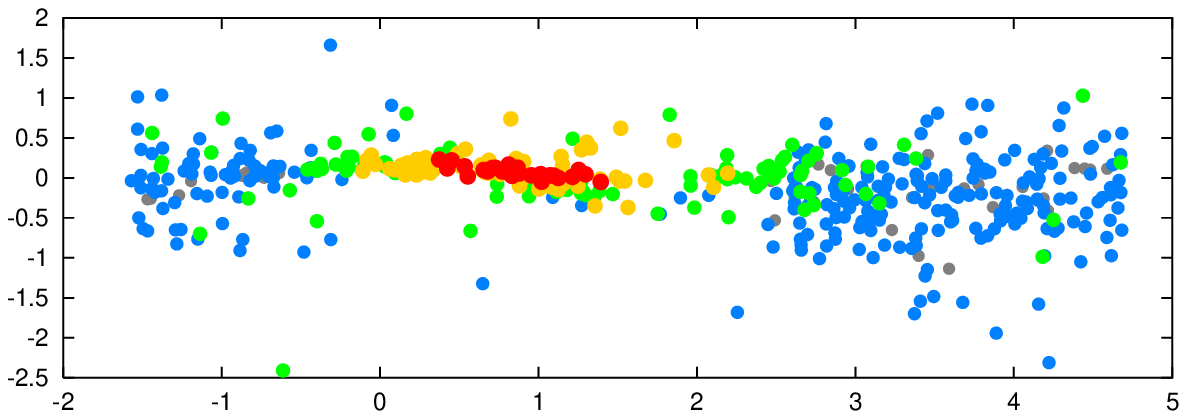}&\raisebox{1cm}{8 hr.}\\

\includegraphics[width=5cm]{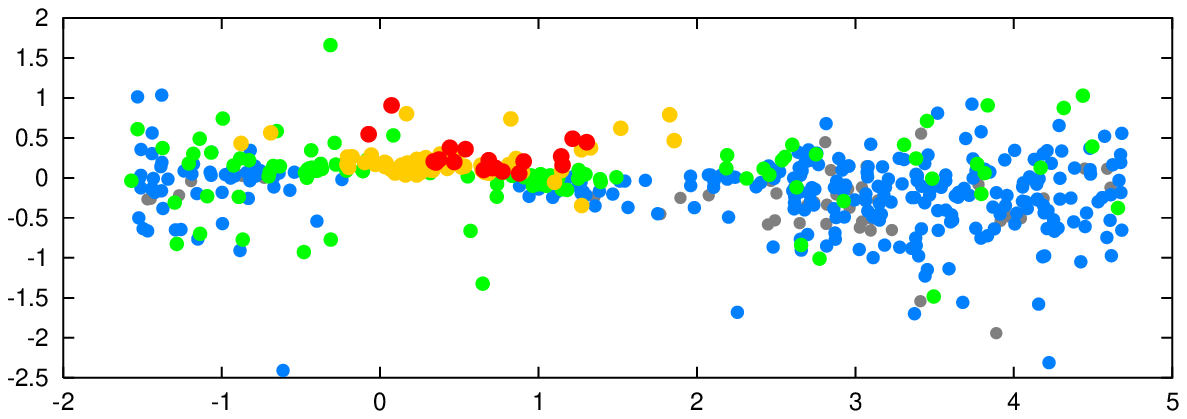}&\raisebox{1cm}{12 hr.}\\

\includegraphics[width=5cm]{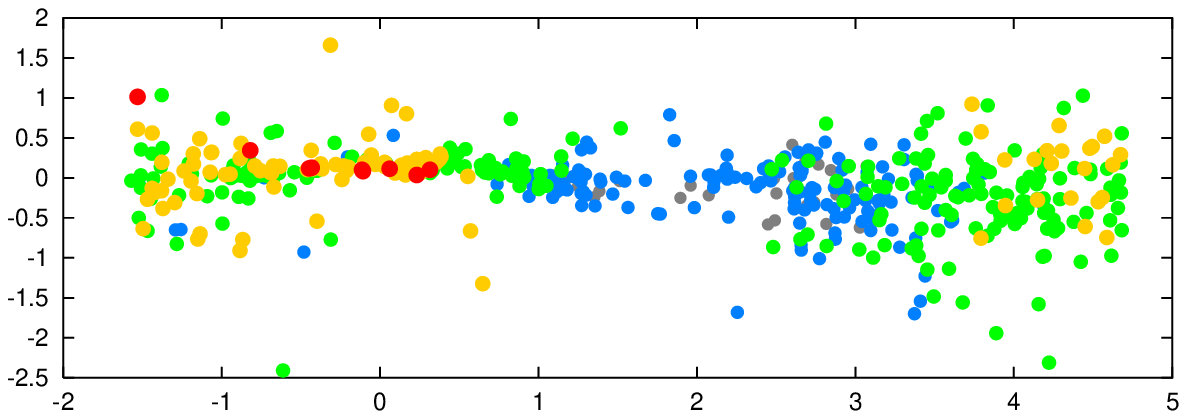}&\raisebox{1cm}{16 hr.}\\

\includegraphics[width=5cm]{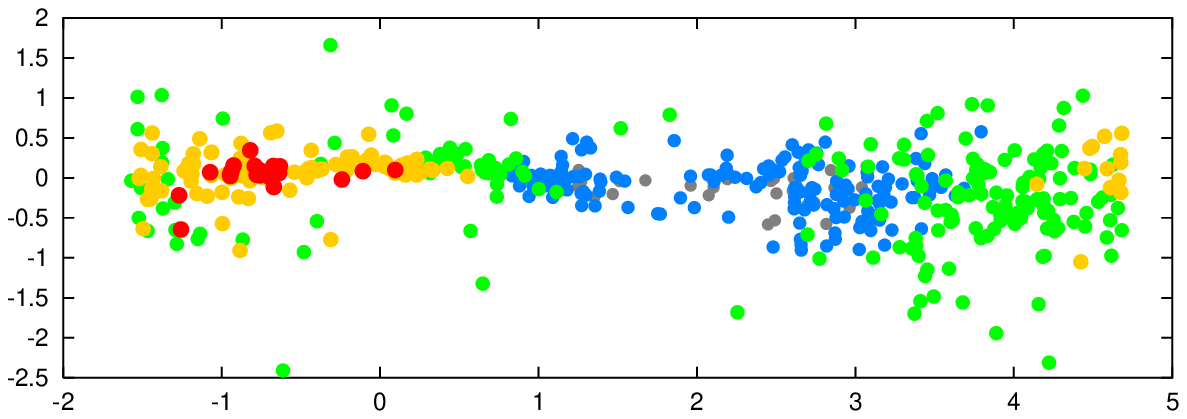}&\raisebox{1cm}{20 hr.}\\

\includegraphics[width=5cm]{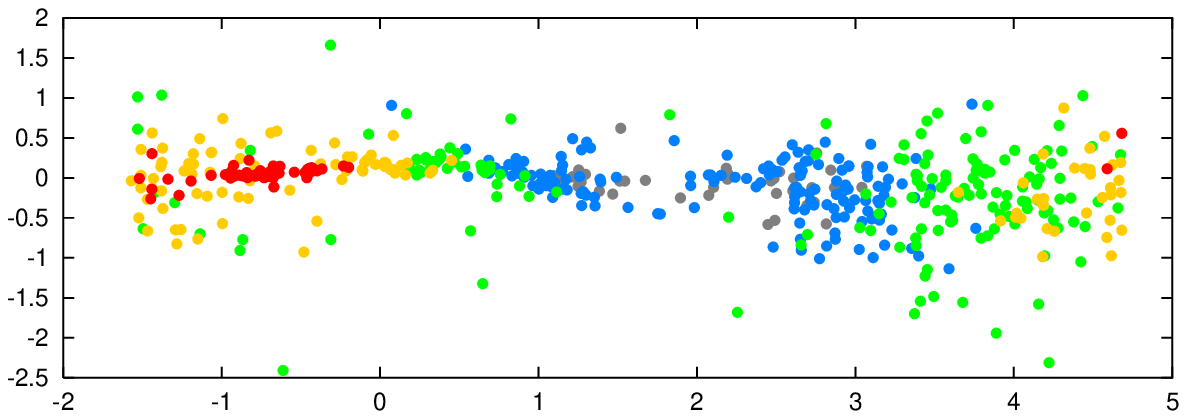}&\raisebox{1cm}{24 hr.}\\
Angle along ring-like structure [rad]  $\rightarrow$
\end{tabular}

\end{document}